\documentclass[pre,twocolumn,showpacs,amsmath,amssymb,floatfix,superscriptaddress]{revtex4}
\usepackage{graphicx}
\usepackage{dcolumn}
\usepackage{bm}
\usepackage{latexsym}
\usepackage{psfrag}
\usepackage{subfigure}
\usepackage{here}
\usepackage{longtable,shapepar}
\renewcommand{\l}{\textrm{L}}
\renewcommand{\r}{\textrm{R}}
\renewcommand{\d}{\textrm{d}}
\newcommand{\w}{\textrm{w}}

\newcommand{\avg}[1]{\left\langle #1\right\rangle}

\begin{document}

\title{Bottleneck-induced transitions in a minimal model for intracellular transport}

\author{Paolo Pierobon} \affiliation{Arnold Sommerfeld Center for
  Theoretical Physics and Center of Nano Science, Department of Physics,
  Ludwig-Maximilians-Universit\"at M\"unchen, Theresienstrasse 37,
  D-80333 M\"unchen, Germany}

\author{Mauro Mobilia} 
 \affiliation{Arnold Sommerfeld Center for
  Theoretical Physics and Center of Nano Science, Department of Physics,
  Ludwig-Maximilians-Universit\"at M\"unchen, Theresienstrasse 37,
  D-80333 M\"unchen, Germany}

\author{Roger Kouyos} \affiliation{Theoretische Biologie,
  Eidegen\"ossische Technische Hochschule Z\"urich,
  Universit\"atstrasse 16, CH-8092 Z\"urich, Switzerland}

\author {Erwin Frey}
 \affiliation{Arnold Sommerfeld Center for
  Theoretical Physics and Center of Nano Science, Department of Physics,
  Ludwig-Maximilians-Universit\"at M\"unchen, Theresienstrasse 37,
  D-80333 M\"unchen, Germany}

\begin{abstract}
  We consider the influence of disorder on the non-equilibrium steady
  state of a minimal model for intracellular transport. In this model
  particles move unidirectionally according to the \emph{totally
    asymmetric exclusion process} (TASEP) and are coupled to a bulk
  reservoir by \emph{Langmuir kinetics}.  Our discussion focuses on
  localized point defects acting as a bottleneck for the particle
  transport. Combining analytic methods and numerical simulations, we
  identify a rich phase behavior as a function of the defect strength.
  Our analytical approach relies on an effective mean-field theory
  obtained by splitting the lattice into two subsystems, which are
  effectively connected exploiting the local current conservation.
  Introducing the key concept of a carrying capacity, the maximal
  current which can flow through the bulk of the system (including the
  defect), we discriminate between the cases where the defect is
  irrelevant and those where it acts as a bottleneck and induces
  various novel phases (called {\it bottleneck phases}).  Contrary to
  the simple TASEP in the presence of inhomogeneities, many scenarios
  emerge and translate into rich underlying phase-diagrams, the
  topological properties of which are discussed.
\end{abstract}


\date{\today}

\maketitle

\section{Introduction}
The effects of disorder on the phase behavior has been investigated in
a multitude of statistical mechanics models and there is by now a good
understanding of the ensuing equilibrium phenomena
\cite{b:ziman:79,b:stinchcombe:83}.  In contrast, in non-equilibrium
statistical mechanics, the effect of disorder on dynamics and
non-equilibrium steady state is far from being well understood
\cite{stinchcombe:02}.

One of the best studied cases is the \emph{totally asymmetric simple
  exclusion process} (TASEP)
\cite{cshaok-vicsek:94,krug-ferrari:96,krug:00, janowsky-lebowitz:92,
  tripathy-barma:97, evans:97, kolomeisky:98, kolwankar-punnoose:00,
  ha-timonen-denijs:03, mirin-kolomeisky:03,shaw-kolomeisky-lee:03,
  shaw-sethna-lee:04, chou-lakatos:04, harris-stinchcombe:04,
  enaud-derrida:04, evans-hanney-kafri:04, juhasz-santen-igloi:05}.
It is defined as a one-dimensional lattice gas where particles are
hopping stochastically in one direction subject to hard-core repulsion
(for reviews see \cite{derrida:98, b:schuetz:01}). Two types of
disorder have been studied: hopping rates may either depend on the
particle attempting to jump (\emph{particle-wise disorder}
\cite{cshaok-vicsek:94,krug-ferrari:96,krug:00} or each lattice site
may be associated may be associated with a random quenched hopping
rate (\emph{site-wise disorder}) \cite{janowsky-lebowitz:92,
  tripathy-barma:97, evans:97, kolomeisky:98, kolwankar-punnoose:00,
  ha-timonen-denijs:03, mirin-kolomeisky:03,shaw-kolomeisky-lee:03,
  shaw-sethna-lee:04, chou-lakatos:04, harris-stinchcombe:04,
  enaud-derrida:04, evans-hanney-kafri:04, juhasz-santen-igloi:05}.
As intuitively expected, such defects generically induce phase
separations if the particle traffic exceeds a certain threshold
\cite{krug:00}.

Our studies are motivated by an important class of intracellular
transport processes mediated by molecular motors like kinesin, dyneins
or myosins moving along cytoskeletal filaments like microtubules or
F-actin \cite{b:howard:01}. The dynamics of each of these molecular
engines is a complicated stochastic process which we idealize by a
Poisson process with a single rate limiting step only. There is now
convincing evidence that transport along the cytoskeletal filaments is
one-dimensional and binding sites are periodically spaced
\cite{ylzid-selin:05}.  Since each of the binding sites can be
occupied by at most one of the molecular motors, particle exclusion
can play a crucial role. Though there is no clear evidence \emph{in
  vivo}, at the moment, that motor densities are large enough for
particle exclusion to dominate the transport properties, there are
\emph{in vitro} investigations underway studying the transport of
kinesin along microtubules at high volume concentrations \cite{foot1}.
Each molecular track has a finite length, and in general one would
like to allow for the enzyme reservoirs at both ends to have different
densities and/or attachment rate at the left and right end to be
different from each other and different from the hopping rate in the
bulk. All this taken together defines the TASEP, introduced originally
as model for the kinetics of biopolymerisation on nucleic acid
template \cite{macdonald-gibbs-pipkin:68, macdonald-gibbs:69}.  For it
to be a proper minimal model for molecular intracellular transport it
also has to account for the fact that microtubules are embedded in the
cytosol with a reservoir of motors in solution. This allows for motors
to attach from the solution to the molecular track or detach from it
and become part of the reservoir again
\cite{lipowsky-klumpp-nieuwenhuizen:01}. Then one arrives at the TASEP
with Langmuir kinetics (TASEP/LK) introduced in
Ref.~\cite{parmeggiani-franosch-frey:03}, which exhibits phase
separation even in the absence of any defects.

In this paper we would like to study the effect of an isolated defect
on the non-equilibrium steady state of this minimal model for
intracellular transport. We focus on a site-wise disorder which may be
mediated by structural imperfections of the microtubular structure or
proteins associated to the microtubules that change the affinity of
the motors with the track.  There has been evidence that these
microtubule associated proteins might even be responsible for some
diseases connected to motor proteins \cite{ebneth-etal:98,
  goldstein:01}.

\begin{figure}[htbp]
  \begin{center}
    \psfrag{alpha}{$\alpha$}
    \psfrag{beta}{$\beta$}
    \psfrag{a}{}
    \psfrag{i=1}{$1$}
    \psfrag{i=N}{$N$}    
    \psfrag{i=k}{$k$}
    \psfrag{L}{${\cal L}$}
    \psfrag{tau=1}{$r=1$}
    \psfrag{q}{$q<1$}
    \psfrag{omega_A}{$\omega_A$}
    \psfrag{omega_D}{$\omega_D$}
    \includegraphics[width=1\columnwidth]{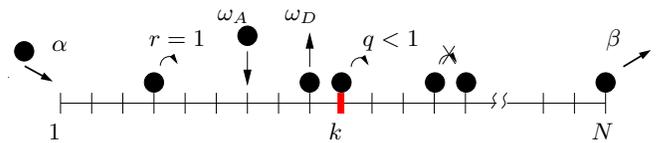}
    \caption{\label{fig:pffdef}Schematic representation of the TASEP
      with on-off kinetics in the presence of a bottleneck at the site
      $i=k$. The allowed moves are: forward jump (with rate $q\neq 1$
      in $i=k$ and $r=1$ elsewhere), entrance at the left boundary
      (with rate $\alpha$) and exit at the right boundary (with rate
      $\beta$), attachment (with rate $\omega_A$), and detachment
      (with rate $\omega_D$) in the bulk.}
    \end{center}
\end{figure}

Let us now introduce the model under investigation. We consider a
simple exclusion process on a finite lattice with $N$ sites (labeled
$i=1,\dots, N$), where the occupation number $n_i$ of each site can be
either $0$ or $1$. The dynamics of the system is described by a fully
unidirectional continuum stochastic process in which each particle
jumps randomly with rate $r_i$ to its right-neighboring vacant site.
At the left boundary particles are introduced in the lattice with rate
$\alpha$, while at the right boundary they are extracted with rate
$\beta$. This defines what is known as Totally Asymmetric Simple
Exclusion Process (TASEP) for $r_i=r$ (we define the time scale by
setting $r=1$). We supplement this process in two ways, as shown in
Fig.~\ref{fig:pffdef}: (a) the system is coupled to a bulk reservoir
via Langmuir kinetics; namely particles can attach to a site in the
bulk with rate $\omega_A$ and can detach with rate $\omega_D$
\cite{foot2}; (b) a special site $k$ of the system represents a
bottleneck and has a slower hopping rate $r_k=q<1$ \cite{foot3}.  Both
cases have been \emph{separately} studied previously and show
non-trivial features. In this work we investigate the combined effect
of these two perturbations to the usual TASEP: our study concerns a
detailed analysis of the non-equilibrium steady state properties, with
emphasis on the resulting phase-diagrams.  Two essential ingredients
to carry on such an analysis are the local density of particles
$\rho_i\equiv\avg{n_i}$ and current profile $j_i\equiv
r_i\avg{n_i\left(1-n_{i+1}\right)}$; where the brackets stand for the
average over the histories. Following the same steps as in
Refs.~\cite{parmeggiani-franosch-frey:03,parmeggiani-franosch-frey:04}
(see also \cite{hinsch05:_from} for a recent review), the stationary
density is shown to obey an hierarchy of equations involving
nearest-neighbor correlation functions:
\begin{eqnarray}
\label{eq:EOMC}
0&=&r_{i-1} \langle {n}_{i-1}(1-{n}_i) \rangle -r_{i}
\langle {n}_{i}(1-{n}_{i+1})\rangle \nonumber\\ &+& \omega_{A} \langle 1-{n}_{i}\rangle - 
\omega_{r} \langle {n}_{i}\rangle, \quad i=2,\dots, N; \\
\label{eq:EOBP}
0&=&\alpha\langle 1-{n}_1 \rangle -
\langle {n}_{1}(1-{n}_{2})\rangle; \\
\label{eq:EOM3}
0&=&\langle {n}_{N-1} (1-{n}_N) \rangle -
\beta \langle {n}_{N}\rangle,
\end{eqnarray} 
where, to account for the defect at site $i=k$, we have:
\begin{eqnarray}
\label{ri}
r_{i}= \begin{cases}
1 & \text{if $i\neq k$}\, ,\\
0<q<1 & \text{if $i=k$}\, .
\end{cases}
\end{eqnarray} 
To be sure to capture an interesting interplay between boundary
induced phenomena and bulk dynamics (see
Refs.~\cite{parmeggiani-franosch-frey:03,
  parmeggiani-franosch-frey:04, pierobon-franosch-frey:06}) we shall
consider a \emph{mesoscopic limit} where local adsorption-desorption
rates are rescaled such that the gross rates $\Omega_{A,D}$ are
comparable to the injection-extraction rates at the boundaries:
\begin{eqnarray} 
  \label{eq:meso}
  \Omega_D=N\omega_D; \quad \Omega_A=N\omega_A\, .
\end{eqnarray}
Keeping fixed the gross rates, in the limit $N\to\infty$, introduces
interesting competition between boundaries and bulk.

This work is organized as follows: the next section summarizes the
main properties of the TASEP, TASEP coupled to the Langmuir kinetics
and TASEP with a single inhomogeneity.  In Section \ref{sec:setup}, we
outline the effective theory on which we build our analysis. Section
\ref{sec:result} is devoted to the discussion of our results
(phase-diagrams and density profiles). Finally, we present our
conclusions in Section \ref{sec:concl}.

\section{Review of previous results: separate role of On-Off kinetics
  and defect}
\label{sec:preliminary}
The effective theory presented hereafter is built on the properties of
models akin to the one under consideration here, namely the TASEP and
TASEP coupled to Langmuir kinetics.  It is thus appropriate to briefly
review the main features of the latter and to summarize the result for
the simple TASEP in the presence of a single inhomogeneity.

\subsection{The  TASEP}
In the absence of attachment and detachment and with uniform hopping
rate, the model defined above reduces to the TASEP.  Much is known
about the non-equilibrium steady state of this paradigmatic model.
Both exact methods \cite{derrida-domany-mukamel:92, derrida-etal:93,
  derrida-evans:93, derrida-evans:99,schuetz-domany:93} and
approximated mean-field solutions
\cite{macdonald-gibbs-pipkin:68,macdonald-gibbs:69,krug:91,kolomeisky-etal:98}
show that as a function of the entrance and exit rates there are three
distinct non-equilibrium steady states. For $\beta<1/2$ and
$\alpha>\beta$ there is a \emph{high density} phase (HD), where both
the density and the current are determined by the exit rate $\beta$.
Mean-field theory gives a spatially constant density $\rho_i=1-\beta$
larger than $1/2$ and a constant current $j_i=\beta(1-\beta)$. Thus
the current is dominated by the low exit rate which acts as a
bottleneck for the transport.

In contrast, for $\alpha<1/2$ and $\alpha<\beta$ the low entrance rate
is the limiting factor for the particle current which is now given by
$j_i=\alpha(1-\alpha)$. Since $\rho_i=\alpha$ is always smaller than
$1/2$ in this parameter range, the phase is also termed the \emph{low
  density} (LD) phase.

If both $\alpha$ and $\beta$ become larger than the critical value
$1/2$ the density becomes constant $\rho^*=1/2$ independently of the
parameters at the boundaries. The current is limited by the particle
exclusion in the bulk and its maximal value is $j^*=1/4$; therefore,
this phase was termed \emph{maximal current} (MC) phase.

\subsection{The TASEP with on-off kinetics}
\label{subsec:pff}
Supplementing the TASEP, a genuine driven (non-equilibrium) process,
 with on-off (or Langmuir, equilibrium) kinetics
results in a system termed TASEP/LK. If the rates of the Langmuir
kinetics are faster than, or comparable to, the hopping rates, the
on-off kinetics always dominates the driven process.  To guarantee the
particles to cover a relevant portion of the lattice before detaching,
the \emph{mesoscopic limit} mentioned in Eq.~(\ref{eq:meso}) has been
introduced \cite{parmeggiani-franosch-frey:03}. This imposes a time scale where 
competition between the boundary, driven processes and the on-off kinetics, 
is effective and results in rich collective phenomena.

Considering Eqs.~(\ref{eq:EOMC})-(\ref{eq:EOM3}) with $r_i=1$, the
mean-field analysis simply consists in neglecting any spatial
correlations resulting in the following decoupling
\emph{approximation}:
\begin{eqnarray}
 \label{eq:MF}
 \avg{n_{i}n_{i+1}} &\approx & \avg{n_{i}}\avg{n_{i+1}}=\rho_i \rho_{i+1}\, ; \\
 \label{eq:cur_MF}
 j_{i}&\approx &\avg{n_{i}}(1-\avg{n_{i+1}})=\rho_i(1-\rho_{i+1})\, .
\end{eqnarray}
Taking the continuum limit with the new spatial variable $0\leq
x\equiv i/N\leq 1$ and using the mesoscopic limit of
Eq.~(\ref{eq:meso}) one obtains:
\begin{eqnarray}
  \label{eq:TASEP_LK_MF}
  (2\rho-1)\partial_x\rho-\Omega_D \rho + \Omega_A (1-\rho)=0\, .
\end{eqnarray}
For the sake of simplicity, let us consider the case where the two
rates are the same ($\Omega_D=\Omega_A=\Omega$). In this case,
Eq.~(\ref{eq:TASEP_LK_MF}) reads:
\begin{eqnarray}
  \label{eq:TASEP_LK_K1}
  (2\rho-1)\left[\partial_x\rho-\Omega\right]=0\, .
\end{eqnarray}
This equation obviously has two solutions: a linear one with slope
$\Omega$ and a constant one coinciding with the critical density of
the TASEP:
\begin{eqnarray}
  \label{eq:sol}
  \rho(x)=\Omega x+C\textrm{\hspace{3mm} and \hspace{3mm}}\rho(x)=\rho^*=\frac 1 2\, ,
\end{eqnarray}
where $C$ is a constant to be determined by the boundary conditions.
The linear density profile results in a space-dependent current, which
in mean-field reads
\begin{eqnarray}
  \label{eq:currden}
  j[\rho(x)]=\rho(x)\left[1-\rho(x)\right]\, .
\end{eqnarray}

\begin{figure}[htbp]
  \begin{center}
    \psfrag{r}[][][1][-90]{$\rho$}
    \psfrag{j}[][][1][-90]{$j$}
    \psfrag{x}{$x$}
    \psfrag{(a)}{(a)}
    \psfrag{(b)}{(b)}
    \psfrag{(c)}{(c)}    
    \psfrag{beta}{$\beta$}
    \psfrag{alpha}{$\alpha$}
    \psfrag{a*}[l]{$\alpha^*=1/2$}
    \psfrag{b*}[c]{$\beta^*=1/2$}
    \psfrag{LD}[][][0.8]{LD}
    \psfrag{HD}[][][0.8]{HD}
    \psfrag{LD/HD}[][][0.8]{LD/HD}
    \psfrag{MC/HD}[][][0.8]{MC/HD}
    \psfrag{LD/MC}[][][0.8]{LD/MC}
    \psfrag{MC}[][][0.8]{MC}
    \psfrag{LD/MC/HD}[][][0.8]{LD/MC/HD}
    \psfrag{X0}{$x_{\w}=0$}
    \psfrag{X1}{$x_{\w}=1$}
    \psfrag{0}[][bl]{0}
    \psfrag{1}[][bl]{1}
    \psfrag{ 0}[][bl]{0}
    \psfrag{ 1}[][bl]{1}
    \psfrag{ 0.2}[][bl]{0.2}
    \psfrag{xw}{$x_{\w}$}
    \psfrag{ra}{$\rho_\alpha$}
    \psfrag{rb}{$\rho_\beta$}
    \psfrag{ja}{$j_\alpha$}
    \psfrag{jb}{$j_\beta$}    
    \includegraphics[width=1\columnwidth]{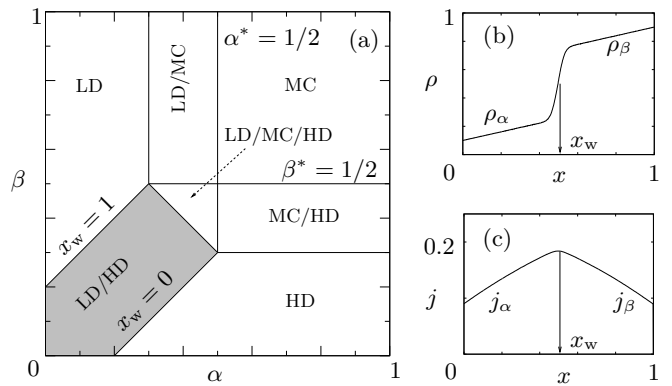}
    \caption{\label{fig:taseplk} (a) Phase diagram of TASEP/LK for
      $K=1$.  One recognizes seven phases: in addition to the TASEP
      LD, HD and MC phases; there are four more coexistence phases, namely
      the LD/HD, LD/MC,
      MC/HD and LD/MC/HD phases.  The shaded region highlights the LD/HD
      coexistence where a localized domain wall appears. (b) Typical density
      profile in the LD/HD phase and (c) the corresponding current profile. At
      the matching point $x_{\w}$ between the left ($j_\alpha$) and
      right ($j_\beta$) currents a domain wall develops and connects the
      left ($\rho_\alpha$) and right ($\rho_\beta$) density profile.}
    \end{center}
\end{figure}

Equation (\ref{eq:TASEP_LK_K1}) has to be supplemented by the boundary
conditions
\begin{eqnarray}
  \label{eq:TASEP_LK_BC}
  \rho(0)=\alpha; \quad \rho(1)=1-\beta\, .
\end{eqnarray}
When the solution of Eq.~(\ref{eq:sol}) cannot be matched continuously
with the left and right boundaries (\ref{eq:TASEP_LK_BC}), the density profile displays a localized 
discontinuity (or shock) in the bulk.  This translates into the emergence of mixed phases.
 The latter are discussed in
Refs.~\cite{parmeggiani-franosch-frey:03,
  parmeggiani-franosch-frey:04} and are summarized in the TASEP/LK phase-diagram of
Fig.~\ref{fig:taseplk}a. Let us illustrate this concept by considering
the transition from the LD to the LD-HD phase. In the former, the
density is determined by the left boundary and reads
\begin{eqnarray}
  \rho(x)=\rho_\alpha(x)\equiv\Omega x+ \alpha\, .
\end{eqnarray}
Lowering the exit rate $\beta$ (with fixed $\alpha$), there is a site
where both the currents imposed by the left and the right boundaries,
meet at $x_{\w}$ (see Fig.~\ref{fig:taseplk}c). This is a mixed LD and HD
phase.  As shown in Fig.~\ref{fig:taseplk}b, the density has a sharp
phase boundary (shock) at $x_{\w}$:
\begin{eqnarray}
  \rho(x)= \begin{cases}
    \label{TASEP_LK_sol_1}
    \rho_\alpha(x)\equiv\Omega x+ \alpha&\textrm{ for }0<x<x_{\w} \\
    \label{TASEP_LK_sol_3}
    \rho_\beta(x)\equiv 1-\beta+\Omega(x-1)&\textrm{ for }x_{\w}<x<1
  \end{cases} .
\end{eqnarray} 

Another feature of the TASEP/LK is the particle-hole symmetry, i.e.\
the properties of the system are invariant under the exchanges
$\alpha\leftrightarrow\beta$, $x\leftrightarrow 1-x$, and
$\rho=1-\rho$. This translate in Fig.~\ref{fig:taseplk} which is
symmetric w.\ r.\ t.\ the line $\alpha=\beta$
\cite{parmeggiani-franosch-frey:03}.

The case $\Omega_A\neq \Omega_D$ follows along the same lines but is
mathematically more involved; for its treatment we refer to
Ref.~\cite{parmeggiani-franosch-frey:04}.  In the following we will
consider $\Omega_A=\Omega_D=\Omega$ and present the case
$\Omega_A\neq\Omega_D$ only at the end of Section \ref{sec:result}.

\subsection{The TASEP with a single inhomogeneity: a brief review}
Before presenting the details of our theoretical treatment of the
TASEP/LK system in the presence of a bottleneck, and to gain some
intuitive understanding of the underlying physics, it is convenient to
outline the properties of the simple TASEP perturbed by a localized
inhomogeneity \cite{janowsky-lebowitz:92,kolomeisky:98}. Here we
consider the TASEP with a defect at site $k$ where the hopping rate is
$q<1$.

\begin{figure}[!ht]
  \psfrag{x}{$x$} \psfrag{r}[][][1][-90]{\hspace{5mm}$\rho$}
   \begin{tabular}{lr}
     \psfrag{a}{(a)}
     \hspace{-5mm}\includegraphics[width=0.55\columnwidth]{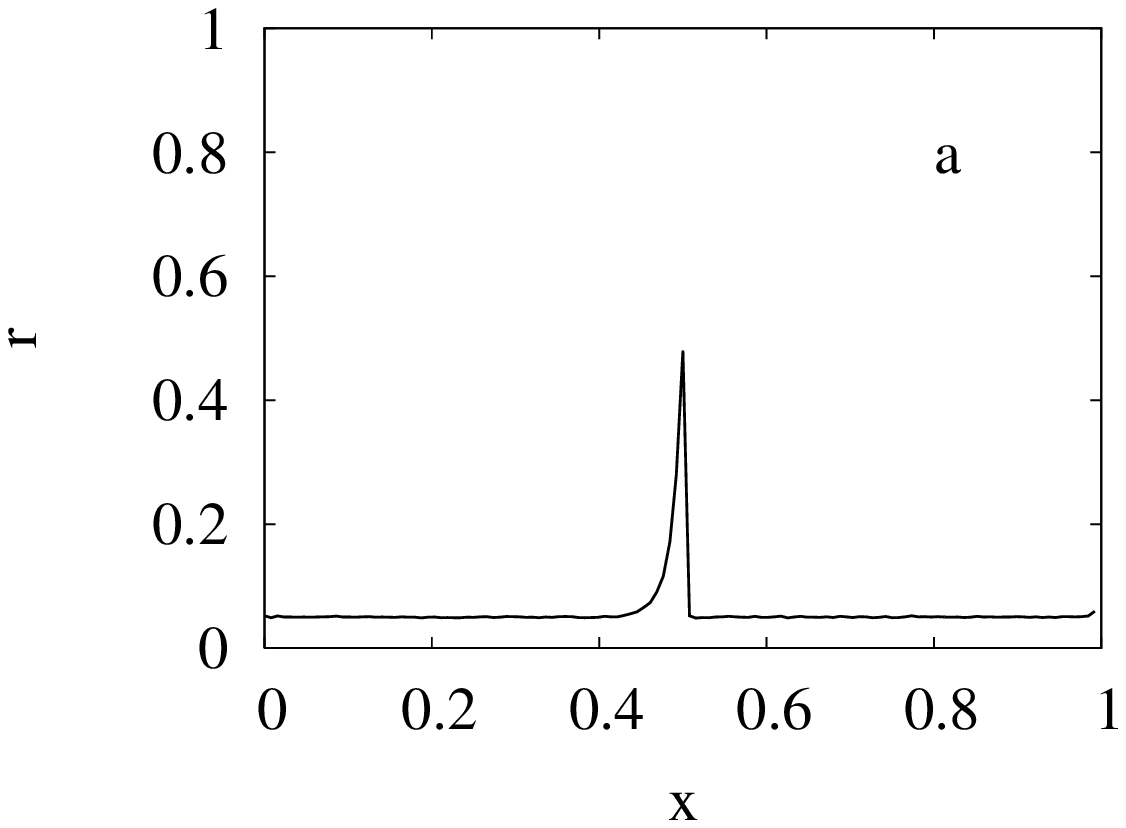}&
     \psfrag{c}{(b)}
     \hspace{-5mm}\includegraphics[width=0.55\columnwidth]{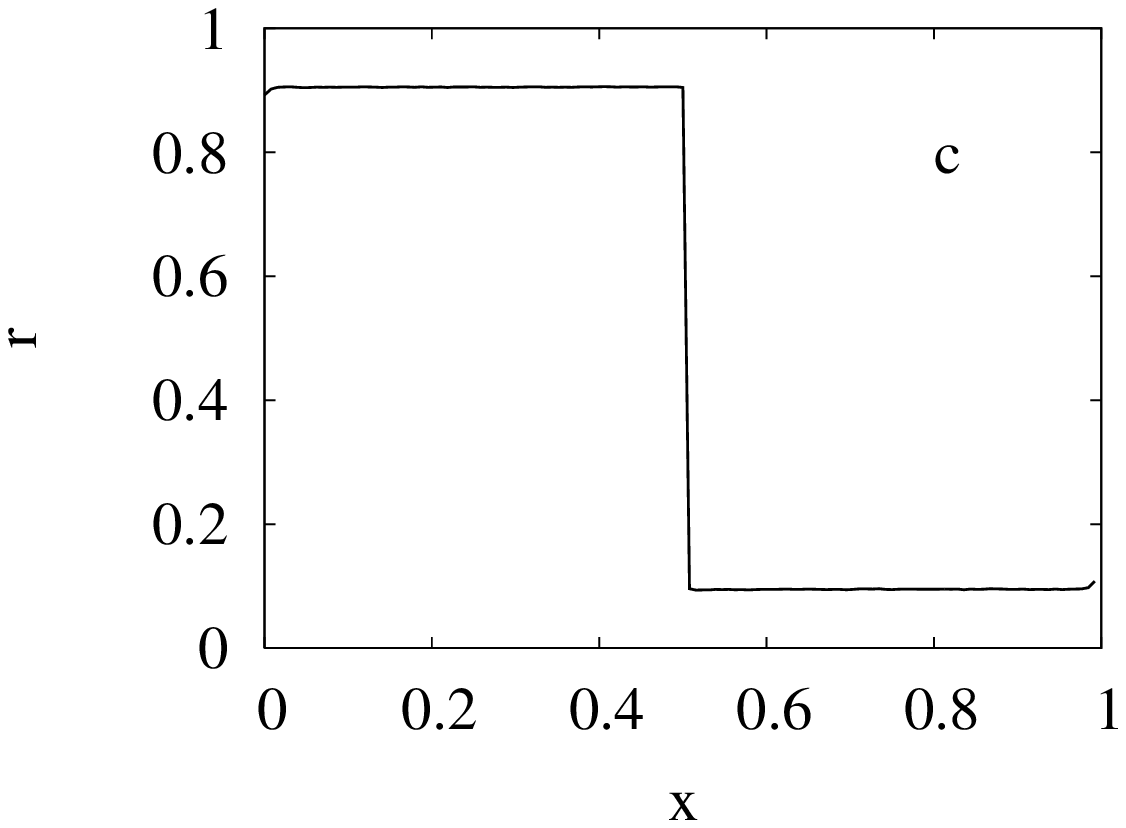}
   \end{tabular}
   \caption{\label{fig:spike1} \quad Transition between a spike (a)
     and a step (b) in the density profile for the TASEP with defect.
     Stochastic simulations are performed on systems of size $N=128$
     with parameters $q=0.1$, $x_{\d}=0.5$, (a)
     $(\alpha,\beta)=(0.05,0.8)$, (b) $(\alpha,\beta)=(0.8,0.8)$.}
\end{figure}

Consider the LD phase, where the system is `diluted' and the particles
are well separated.  In such a ``low traffic'' situation, one does not
expect any macroscopic effects arising from the presence of the
bottleneck. One rather expects a local peak in the density profile
(see Fig.~\ref{fig:spike1}a): $\rho_{i\neq k}=\alpha$ and
$\rho_{k}=\alpha + h$ (LD phase), where $h$ is the height of the local
jump imposed by the defect. Since an exclusion process without
coupling to a bulk reservoir has a spatially constant current
$j_i=\alpha(1-\alpha)$, one finds $h=h(\alpha, q)=\alpha(1-q)/q$. Thus
the height of the density peak increases with $\alpha$ and $q^{-1}$.
This can certainly not happen without bound. In fact, $h$ cannot exceed
$h_{\max} =1-2\alpha$. This may be seen as follows. If $h>h_{{\rm
    max}}$, there would be a site $i_1$ such that $\rho_{i_1}<\alpha +
h_{\max}=1-\alpha$ and $\rho_{i_1 +1}>\alpha + h_{\max}$.  This case
has to be discarded as it contradicts current conservation:
$j_{i_1}=\rho_{i_1}(1-\rho_{i_1})<j_i=\alpha(1-\alpha)$.  Therefore,
for given $\alpha$ and $\beta$, there is a critical value of
$q_{\alpha}^{*}$ below which the local peak is no longer a possible
solution. The critical $q_{\alpha}^{*}$ follows from the requirement that the
height of the peak cannot exceed $h_{\max}$, i.e.\ $h(\alpha,q_{\alpha}^{*})=h_{\max}$. One thus finds
\begin{eqnarray}
 q_{\alpha}^{*}=\frac{\alpha}{1-\alpha}\, .
\end{eqnarray}
Equivalently, if one keeps $q$ fixed, a local peak is obtained for
\begin{eqnarray}
 \alpha\leq \frac{q}{1+q}=\alpha_c\, .
\end{eqnarray}

It directly follows from the underlying particle-hole symmetry that
the same reasoning holds in the HD phase ($\alpha >\beta$ and
$\beta<1/2$). There the system is so `packed' that the bottleneck is
only responsible for a local dip in the density profile: $\rho_{i\neq
  k}=1-\beta$ and $\rho_{k}=1-\beta-h(\beta,q)$, with
$h(\beta,q_{\beta}^{*})<h_{\max}'=1-2\beta$. Thus, in the HD phase, a
local dip is only possible for 
\begin{eqnarray}
  q_{\beta}^{*}\equiv\frac{\beta}{1-\beta}<q
\end{eqnarray}
or, keeping $q$ fixed, for 
\begin{eqnarray}
  \beta\leq \frac{q}{1+q}=\beta_c\, .
\end{eqnarray}

We conclude that, for a fixed value of $q$, there are critical values
for the entrance and exit rates above which the defect leads to a jump
in the density profile. In contrast to the local density perturbation
(``spike'') for small $\alpha$ and $\beta$, this is a macroscopic
effect and does not vanish in the thermodynamics limit (while the
spike width scales as $1/N$).

Denoting respectively by $\rho^{\l}$ and
$\rho^{\r}$ the densities on the left and right of the inhomogeneity, one
expects a sharp step profile displaying a jump at the defect position: 
\begin{eqnarray}
  \rho^{\l}=\frac 1 2+\delta\textrm{\quad and\quad }\rho^{\r}=\frac 1
2-\delta\ .  
\end{eqnarray}
Thus, the current through the defect is
\begin{eqnarray}
  j_k=q\rho_{k}(1-\rho_{k+1})=q\left[\frac 1 2+\delta\right]\left[1-\left(\frac 1 2-\delta\right)\right]\, ,
\end{eqnarray}
while the current flowing through the bulk of the track reads
\begin{eqnarray}
  j_{i\neq k}=\left(\frac 1 2+\delta\right) \left(\frac 1 2-\delta\right)\, .
\end{eqnarray}
As the current is conserved, these expressions should coincide, which
gives the jump in the density profile: 
\begin{eqnarray}
  2 \delta=\frac{1-q}{1+q}\, .
\end{eqnarray}
This is an important quantity which measures the \emph{strength of the
  defect}: $\delta=0$ when $q=1$ and $\delta=1/2$ for $q=0$.

Thus, for the density profile, one finds
\begin{eqnarray}
  \rho^{\l}=\frac{1}{1+q}\textrm{\hspace{3mm} and\hspace{3mm} }\rho^{\r}=\frac{q}{q+1}\, ,
\end{eqnarray}
while the $q-$dependent current flowing through the entire system,
playing the role of an effective (bottleneck induced) MC, reads
\begin{eqnarray} 
  \label{eq:jdef}
  j_d^*=\frac{q}{(1+q)^2}<\frac 1 4\, .
\end{eqnarray}
Since there is a jump and the density profile is flat on both sides of the defect,
one may effectively split the systems into two parts connected by
current conservation \cite{kolomeisky:98}; (see Fig.~\ref{PFF_def}).
The effective exit rates of the left subsystem and the entrance of the
right one are therefore:
\begin{eqnarray}
  \label{eq:effrate}
  \alpha_{\rm eff}=\beta_{\rm eff}=\frac{q}{1+q}\, .
\end{eqnarray}

It follows from this discussion that for the TASEP only the MC phase
is affected by the presence of a bottleneck, while the LD and HD
phases remain unaltered. The resulting phase-diagram therefore still
displays the same topological features (with three phases: LD, HD and
MC) as in the homogeneous case, with the lines $\alpha=\beta=1/2$,
delimiting the MC phase, lowered to by $\alpha=\alpha_c$ and
$\beta=\beta_c$.

\section{Effective mean-field theory}
\label{sec:setup}
In this section, we describe an effective mean-field theory for the
non-equilibrium steady state of the TASEP/LK in the presence of a
bottleneck.

\begin{figure}[!ht]
  \psfrag{x}{$x$} \psfrag{r}[][][1][-90]{\hspace{5mm}$\rho$}
   \begin{tabular}{lr}
     \psfrag{b}{(a)}
     \hspace{-5mm}\includegraphics[width=0.55\columnwidth]{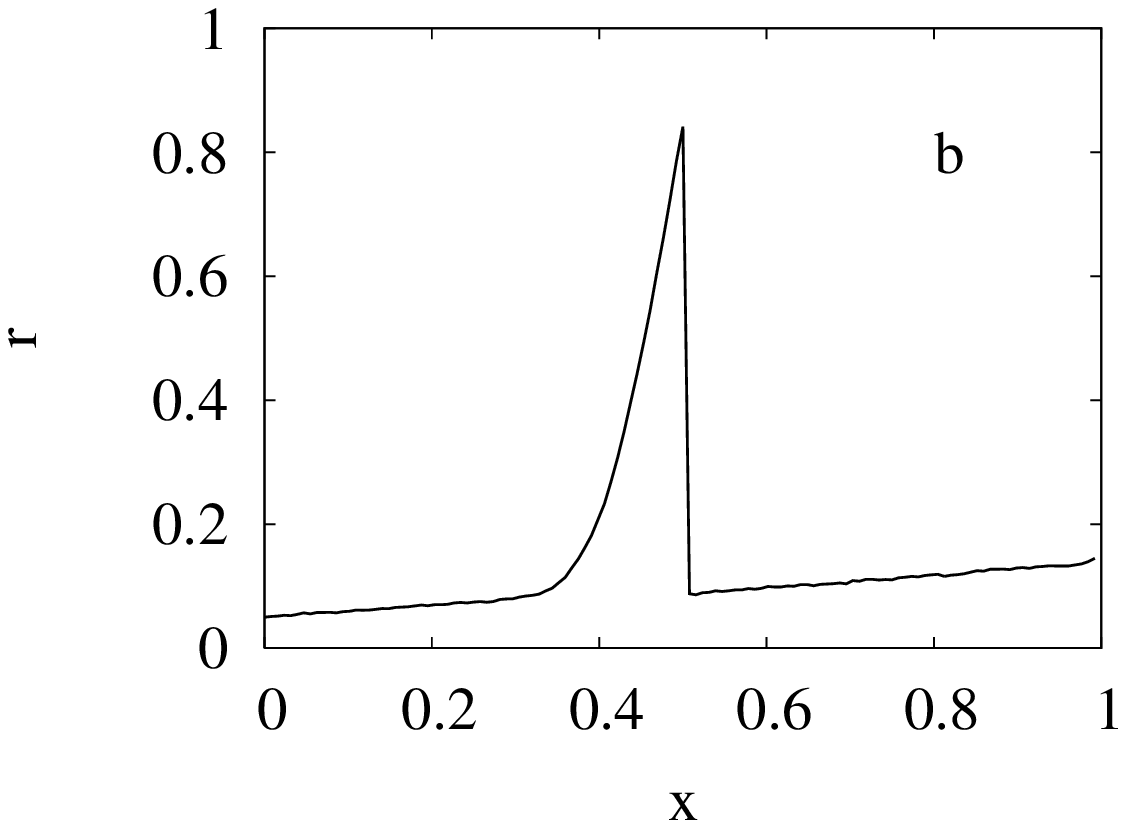}&
     \psfrag{d}{(b)}
     \hspace{-5mm}\includegraphics[width=0.55\columnwidth]{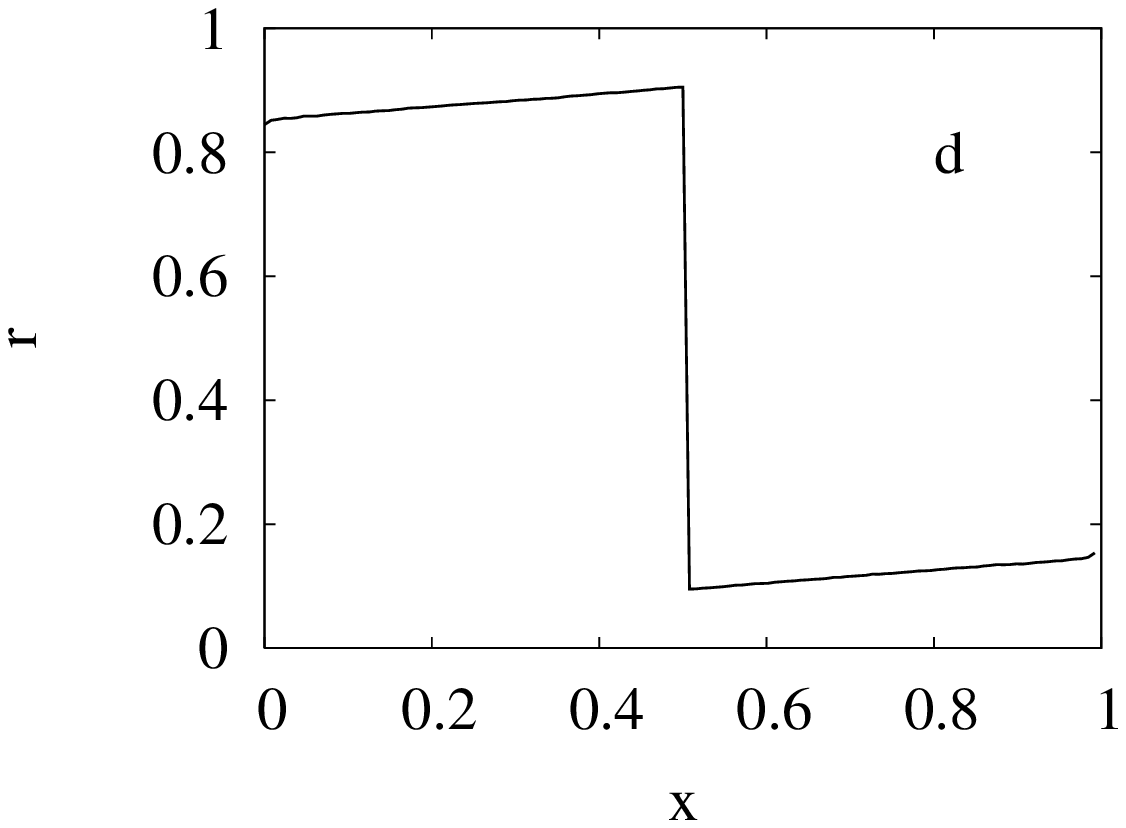}
   \end{tabular}
   \caption{\label{fig:spike2} \quad Transition between spike (a) and
     step (b) for the TASEP/LK with defect. Stochastic simulations are
     performed on systems of size  $N=128$ with parameters $q=0.1$,
     $x_{\d}=0.5$, $\Omega=0.1$, (a) $(\alpha,\beta)=(0.05,0.8)$, (b)
     $(\alpha,\beta)=(0.8,0.8)$.}
\end{figure}

\begin{figure}[h!]
  \begin{center}
    \psfrag{alpha}{$\alpha$} \psfrag{beta}{$\beta$}
    \psfrag{alphaeff}{$\alpha_{\textrm{eff}}$}
    \psfrag{betaeff}{$\beta_{\textrm{eff}}$}
    \psfrag{ralpha}[c]{$\rho_1=\alpha$}
    \psfrag{rbeta}[c]{$\rho_N=1-\beta$}
    \psfrag{ralpha1}[c]{$\rho(0)=\alpha$}
    \psfrag{rbeta1}[c]{$\rho(1)=1-\beta$} \psfrag{q1}[l]{$\rho_k$}
    \psfrag{q2}[l]{$\rho_{k+1}$} \psfrag{q1a}[c]{$\frac{1}{1-q}$}
    \psfrag{q2a}[c]{$\frac{q}{1-q}$} \psfrag{q}{$q$} \psfrag{x=k}{$i=k$}
    \psfrag{I}{L}
    \psfrag{II}{R}
    \vspace{5mm}
    \includegraphics[width=0.85\columnwidth]{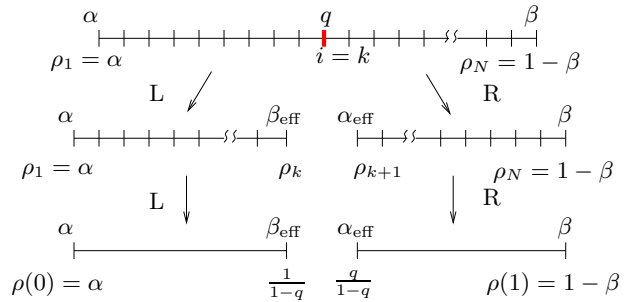}
    \caption{\label{PFF_def}Schematic representation of the division
      into two subsystems allowing to apply a mean-field theory. The last step
      (bottom of the figure), illustrates the continuum limit that we
      are considering (see the main text).}
        \end{center}
\end{figure}

As for the TASEP perturbed by the presence of a localized defect,
there are regions of the parameter space where the inhomogeneity
affects the system only locally.  In other regions, the defect has
macroscopic effects and is said to be relevant (see
Fig.~\ref{fig:spike1}).  In this case, the properties of the system
are studied by splitting the lattice at the defect site $k$ into two
subsystems (again termed ${\l}$ and ${\r}$) and performing a continuum
limit (see Fig.~\ref{PFF_def}). In such a limit, the position of the
defect becomes $x_{\d}\equiv \lim_{N\to\infty}k/N$, and the density
can be written as
\begin{eqnarray}
  \label{eq:pffcases}
  \rho(x)= \begin{cases}
    \label{rho_l}
    \rho^{\l}(x)\quad ,0\leq x\leq x_{\d}&\\
    \label{rho_r}
    \rho^{\r}(x)\quad ,x_{\d}< x\leq 1
  \end{cases} \, .
\end{eqnarray} 

In the presence of a bottleneck, the current is \emph{locally}
conserved for the same reasons as in the TASEP/LK system
\cite{parmeggiani-franosch-frey:04} [essentially because the
attachment/detachment rates scale as in Eq.~(\ref{eq:meso})]:
\begin{eqnarray}
  \label{cons}
  j_{i+1}-j_{i}=\frac{1}{N}\left[\Omega_A -(\Omega_A + \Omega_D) \rho_{i+1}\right] \xrightarrow[N\rightarrow \infty]{} 0\, .
\end{eqnarray}
This allows to couple both subsystems R and L along the same lines as
for the simple TASEP perturbed by an isolated inhomogeneity, with
effective right and left boundaries given by Eq.~(\ref{eq:effrate}).

Suppose now that we are in the parameter range where the defect
dominates the current and density profile. Then we may argue as in the
previous section and obtain for the current at the defect site :
\begin{eqnarray}
\label{jk}
  j_{k}=j_{k+1}=\frac{q}{(1+q)^2}\, .
\end{eqnarray}
and the corresponding densities
\begin{eqnarray}
\label{rhok}
  \rho_{k}=\frac{1}{1+q}\, , \quad  \rho_{k+1}=\frac{q}{1+q}\, .
\end{eqnarray}
Again this implies a defect induced jump in the density of magnitude
\begin{eqnarray}
\label{delta}
 2\delta\equiv \rho_{k}-\rho_{k+1}=\frac{1-q}{1+q}\, .
\end{eqnarray}
A key difference to our previous discussion of a pure TASEP in
Sec.\ref{sec:preliminary} is that these properties apply only locally
in the vicinity of the defect. In the continuum limit, the density
profile is linear [see Eq.~(\ref{eq:sol})] and has to match the
boundary conditions given by Eqs.~(\ref{rhok}). Hence the density
profile $\rho_{\d}(x)$ imposed by the defect reads
\begin{eqnarray}
  \label{rho_d}
  \rho_d(x)= \begin{cases}
    \label{rho_d_l}
    \Omega(x-x_{\d})+\frac{1}{1+q}\quad ({\l})&\\
    \label{rho_d_r}
    \Omega(x-x_{\d})+\frac{q}{1+q}\quad({\r})
  \end{cases}.
\end{eqnarray} 
Since the mean-field current-density relationship is given by
Eq.~(\ref{eq:currden}), this immediately implies for the current
\begin{eqnarray}
\label{jd}
j_{d}= 
\left(\Omega |x - x_{\d}| +\frac{q}{1+q}\right)\left(\frac{1}{1+q} - \Omega |x_{\d}-x|\right)\, .
\end{eqnarray}

\begin{figure}[!ht]
  \psfrag{x}[c]{$x$} \psfrag{j}{$j$}
  \psfrag{1/2}{$\frac 1 2$}
  \psfrag{1/4}{$\frac 1 4$}
  \psfrag{0}{$0$}
  \psfrag{1}{$1$}
  \psfrag{r}{$\rho$}
  \psfrag{j}{$j$}  
  \psfrag{xw-xi}{$x_{\d}-\xi$}
  \psfrag{xw+xi}{$x_{\d}+\xi$}
  \psfrag{xi}{$\xi$}
  \psfrag{xw}{$x_{\d}$}  
  \psfrag{(a)}{(a)}\psfrag{(b)}{(b)}
  \includegraphics[width=0.8\columnwidth]{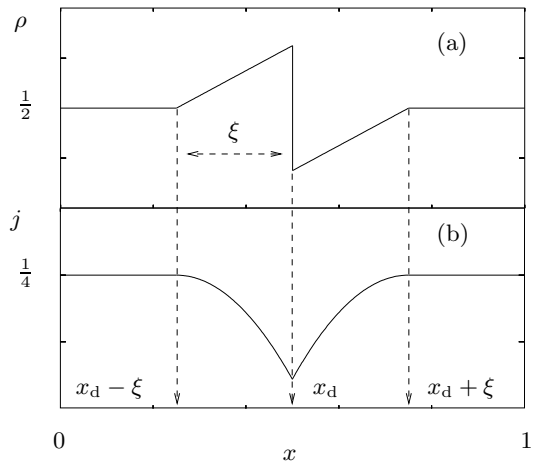}
  \caption{\label{fig:zigzag}\quad Sketch of the signature of the
    strong defect (zigzag shape) in the density (a) and current (b)
    profile.}
\end{figure}

In stark contrast to the simple TASEP, here the current is a
space-dependent quantity. As illustrated in Fig.~\ref{fig:zigzag}, the
signature of a defect is the depletion on a \emph{macroscopic} scale
of the current profile (Fig.~\ref{fig:zigzag}b).  Correspondingly, the
density displays a `zigzag' shape with a jump at $x_{\d}$ and a linear
profile in its vicinity as expressed in Eq.~(\ref{rho_d}) (see
Fig.~\ref{fig:zigzag}a).

At $x=x_{\d} \pm \xi$, the current imposed by the defect reaches the
maximal value $j^*=1/4$.  Thus, the depletion or \emph{screening
  length} $\xi$ induced by the bottleneck is the solution of
$\rho_d(x_{\d}\pm \xi)=\rho^*=1/2$ and reads
\begin{eqnarray}
  \label{xi}
  \xi=\frac{\delta}{\Omega}.
\end{eqnarray}
The defect is thus screened best for a strong coupling to the
reservoir ($\Omega \gg\delta$) and, of course, if the inhomogeneity is
weak, i.e.\ $\delta \to 0$.

Since the screening length increases with the strength of the defect
it can even become larger than the length of the sub-systems.  This
happens when $\xi$ is larger than the distance to either one of the
boundaries.  According to the conditions $x_{\d}-\xi <0$ and $x_{\d}+\xi>1$,
this happens when $\delta>\max(\delta_1,\delta_2)$ (and fixed
$\Omega$) or, equivalently, when $\Omega<\min(\Omega_1,\Omega_2)$
(with fixed $q$), where
\begin{eqnarray}
  \label{critd}
  \delta_1= \Omega x_{\d}&\textrm{\quad and \quad}
  &\delta_2=\Omega(1-x_{\d})\\
  \Omega_1=\frac{\delta}{x_{\d}}&\textrm{\quad and \quad}
  &\Omega_2=\frac{\delta}{1-x_{\d}}\, .
\end{eqnarray}
If $\delta<\min(\delta_1,\delta_2)$ the screening length is shorter
than both lengths of the subsystems.  When the location of the defect
is not centered, two additional cases arise: for
$\delta_1<\delta<\delta_2$ ($\delta_2<\delta<\delta_1$) the screening
length stays within the subsystem R (L), while it is larger than the
size of the sub-lattice L (R).

In the TASEP and in the TASEP/LK the maximal current that can flow
through the system is a constant $j^*=1/4$. For the TASEP with a
defect that value is lowered to $j_d^*=q/(1+q)^2$ [see
Eq.~(\ref{eq:jdef})].  As the current is space dependent and locally
conserved in the TASEP/LK with an inhomogeneity, the maximal flow of
particles through the bulk varies spatially. This suggests to term
such a quantity the \emph{carrying capacity} ${\cal C}(x)$ of the
system.

\begin{figure}[!ht]
  \psfrag{x}[c]{$x$} \psfrag{j}{$j$} \psfrag{j*}{$j^*$}
  \psfrag{xd}{$x_{\d}$} \psfrag{x1}{$x_{\d}-\xi$} \psfrag{x2}{$x_{\d}+\xi$}
  \psfrag{ja}{$j_\alpha$} \psfrag{jb}{$j_{\alpha_{\rm eff}}$}
  \psfrag{jl}{$j_{\beta_{\rm eff}}$} \psfrag{jr}{$j_{\beta_{\rm
        eff}}$}
  \psfrag{(a)}{(a)}\psfrag{(b)}{(b)}\psfrag{(c)}{(c)}\psfrag{(d)}{(d)} 
   \begin{tabular}{lr}
     \hspace{-5mm}
     \psfrag{j}{${\cal
         C}_1(x)$}\includegraphics[width=0.55\columnwidth]{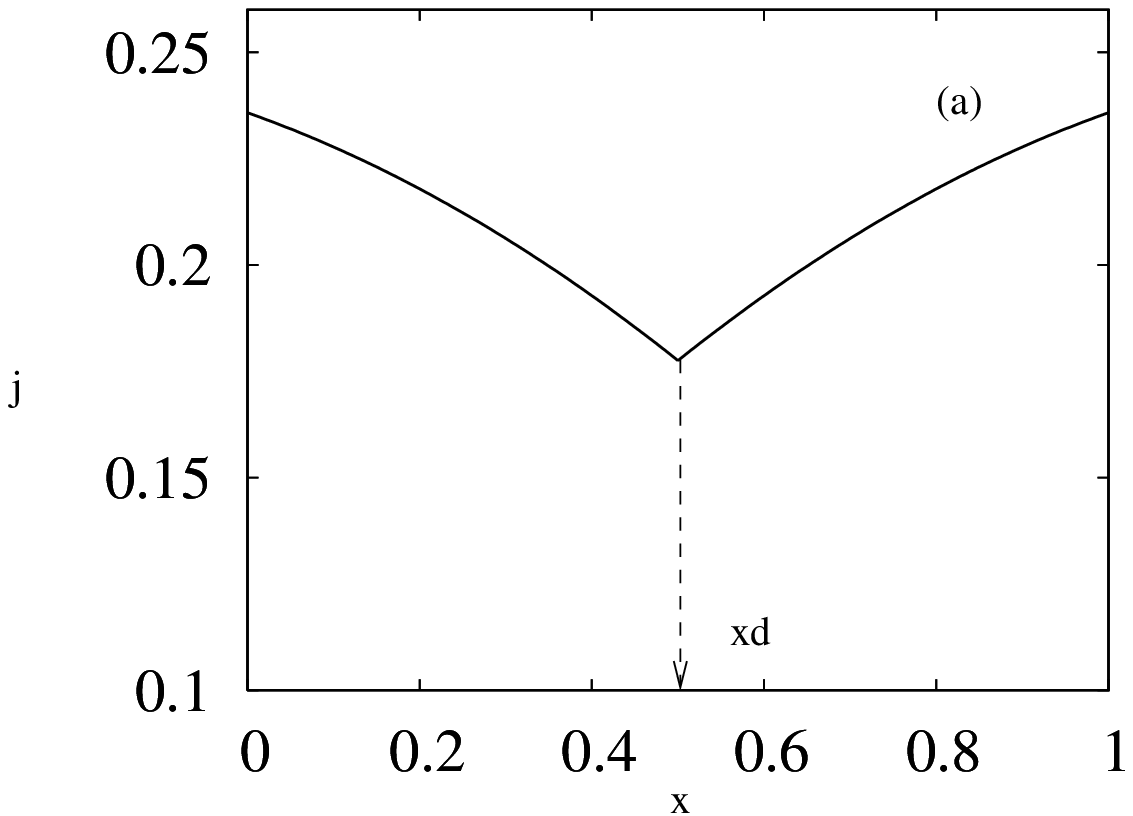}&
     \hspace{-5mm} 
     \psfrag{j}{${\cal
         C}_2(x)$}\includegraphics[width=0.55\columnwidth]{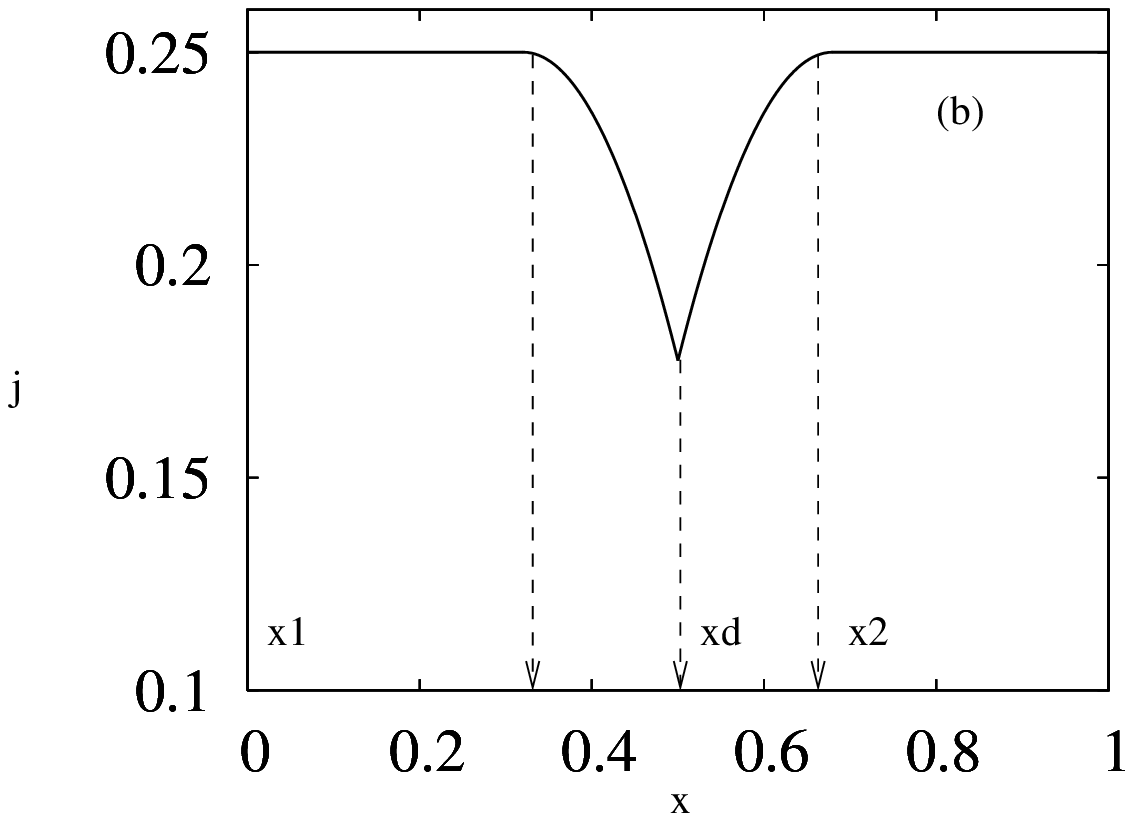}\\
     \hspace{-5mm}  
     \psfrag{j}{${\cal
         C}_3(x)$}\includegraphics[width=0.55\columnwidth]{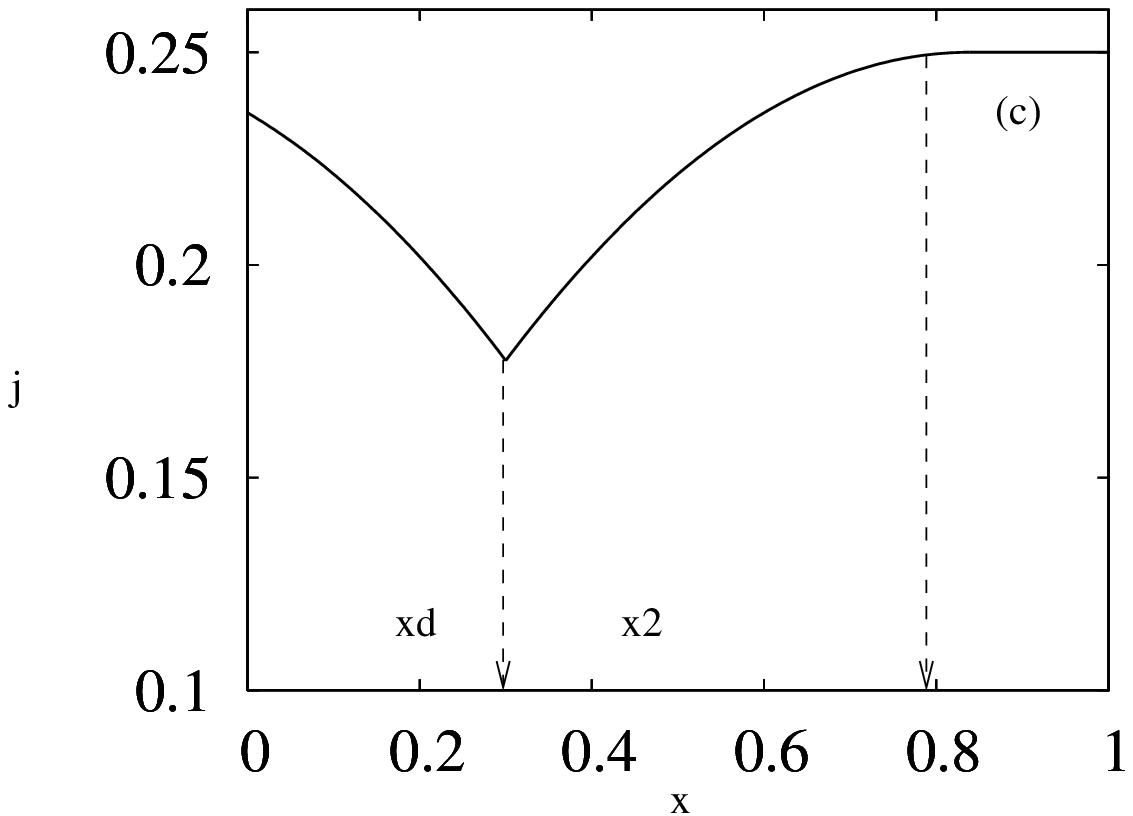}&
     \hspace{-5mm}
     \psfrag{j}{${\cal
         C}_4(x)$}\includegraphics[width=0.55\columnwidth]{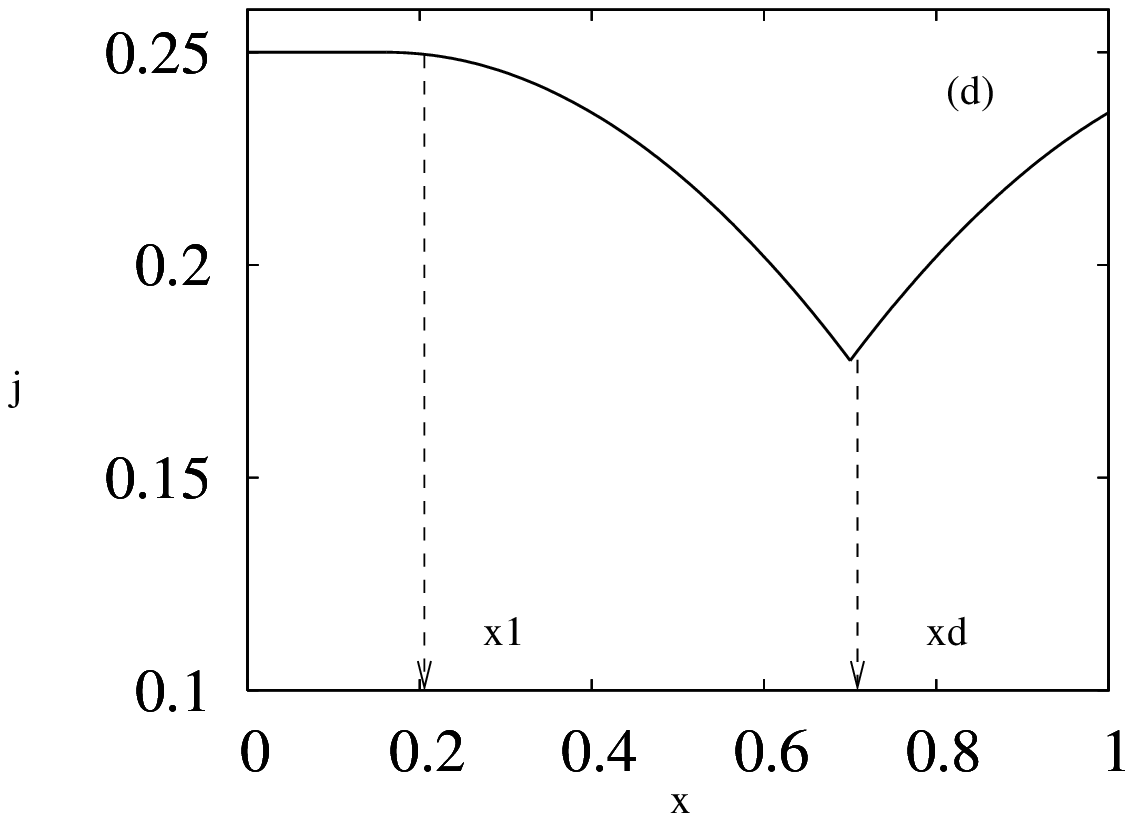}
   \end{tabular}
   \caption{ \label{fig:excurrb}\quad The four typical carrying
     capacity profiles ${\cal C}(x)$ displayed by the system
     (parameters $\Omega=\{0.3,1.5,0.5,0.5\}$,
     $x_{\d}=\{0.5,0.5,0.7,0.3\}$, $q=0.3$).  Depending on the defect
     strength $q$, on the position $x_{\d}$ of the defect and on the rate
     $\Omega$, the defect imposed current $j_d$ combines in four
     different ways with the maximal current $j^*$.  Each of these
     profiles induce topologically distinct phase diagram (see the
     text).}
\end{figure}

The more drastic effect of the bottleneck appears when its strength satisfies
$\delta>\max(\delta_1,\delta_2)$. In this case, defect screens the entire
system and, as shown in Fig.~\ref{fig:excurrb}a, the
carrying capacity reads
\begin{eqnarray}
{\cal C}(x)={\cal C}_1(x)= j_d(x)\, .
\end{eqnarray}

In intermediate cases, when the screening length covers part of the
system, one has three possible scenarios illustrated in
Fig.~\ref{fig:excurrb}b-\ref{fig:excurrb}d, namely
\begin{eqnarray}
  {\cal C}(x)&=&{\cal C}_2(x)=\begin{cases} j_d(x)&, |x-x_{\d}|<\xi\\ j^* &, \text{else}
  \end{cases}.
\end{eqnarray}
\begin{eqnarray}{\cal C}(x)={\cal C}_3(x)=\begin{cases}
    j_d(x)&, 0<x<x_{\d}+\xi\\
    j^* &, x_{\d}+\xi\leq x<1
  \end{cases}.
\end{eqnarray}
\begin{equation}{\cal C}(x)={\cal C}_4(x)=\begin{cases} j^*&,
    0<x<x_{\d}-\xi\\ j_d(x) &, x_{\d}-\xi\leq x<1
  \end{cases}.
\end{equation} 
For each possible carrying capacity of Fig.~\ref{fig:excurrb}, one
obtains different phase diagrams.  All these scenarios are discussed
in the next section.

To determine the global current and density profiles from the carrying
capacity, one compares the latter with the currents imposed by the
open boundaries:
\begin{eqnarray}
\label{ja}
j_{\alpha}&=& \left(\alpha + \Omega x\right) \left(1-\alpha 
  - \Omega x\right)\,,  \\
\label{jb}
j_{\beta}&=&\left[\beta+ \Omega (1- x)\right]
\left[1-\beta - \Omega(1- x)\right],
\end{eqnarray}
corresponding to the left and right density profiles
\begin{eqnarray}
  \label{rho_a}
  \rho_{\alpha}(x)&=& \Omega x+ \alpha, \\
  \label{rho_b}
  \rho_{\beta}(x)&=&\Omega(x-1) + (1-\beta).
\end{eqnarray} 
As ${\cal C}(x)$ acts as an effective maximal current in the bulk,
$j_{\alpha}(x)$ and $j_{\beta}(x)$ cannot exceed its value. Actually,
as for the TASEP (where ${\cal C}=1/4$), the entrance and exit
boundaries only matter on the macroscopic part of the system where
they impose currents smaller than ${\cal C}(x)$. Thus, by matching
$j_{\alpha}(x)$ and $j_{\beta}(x)$ with the carrying capacity, one
determines boundaries separating boundary-induced phases and mixed
ones.

\section{Results: phase-diagram of the TASEP/LK in the presence of a
  bottleneck}
\label{sec:result}
In this section we discuss the four possible scenarios arising for
each carrying capacity presented above. We explicitly construct the
density profiles and the phase diagrams when the screening length is
larger than the size of the two subsystems (carrying capacity ${\cal
  C}_1(x)$) and when $\xi$ is shorter than the size of the two
subsystems (carrying capacity ${\cal C}_2(x)$). In the last subsection
we extend the results obtained so far for the case of equal attachment
and detachment rates ($\Omega_A=\Omega_D$) to the more general
situation $\Omega_A\neq\Omega_D$.

\subsection{Large screening length: the case ${\cal C}_1(x)$}
\begin{figure}[htbp]
  \begin{center}
    \psfrag{x}{$x$}
    \psfrag{j}{$j$}
    \psfrag{case1}[][][1]{$j_\alpha$ Case 1}
    \psfrag{case2}[][][1]{$j_\alpha$ Case 2}
    \psfrag{case3}[][][1]{$j_\alpha$ Case 3}
    \psfrag{Case 1}[][][1]{Case 1}
    \psfrag{Case 2}[][][1]{Case 2}
    \psfrag{Case 3}[][][1]{Case 3}
    \psfrag{ 0}{$0$}
    \psfrag{ 0.1}{$0.1$}
    \psfrag{ 0.2}{$0.2$}
    \psfrag{ 0.4}{$0.4$}
    \psfrag{ 0.6}{$0.6$}    
    \psfrag{ 0.8}{$0.8$}
    \psfrag{ 1}{$1$}
    \psfrag{c1}[][][0.8]{${\cal C}_1(x)$}
    \psfrag{jb}{$j_\beta$}
    \vspace{1.5cm}
    \includegraphics[width=1\columnwidth]{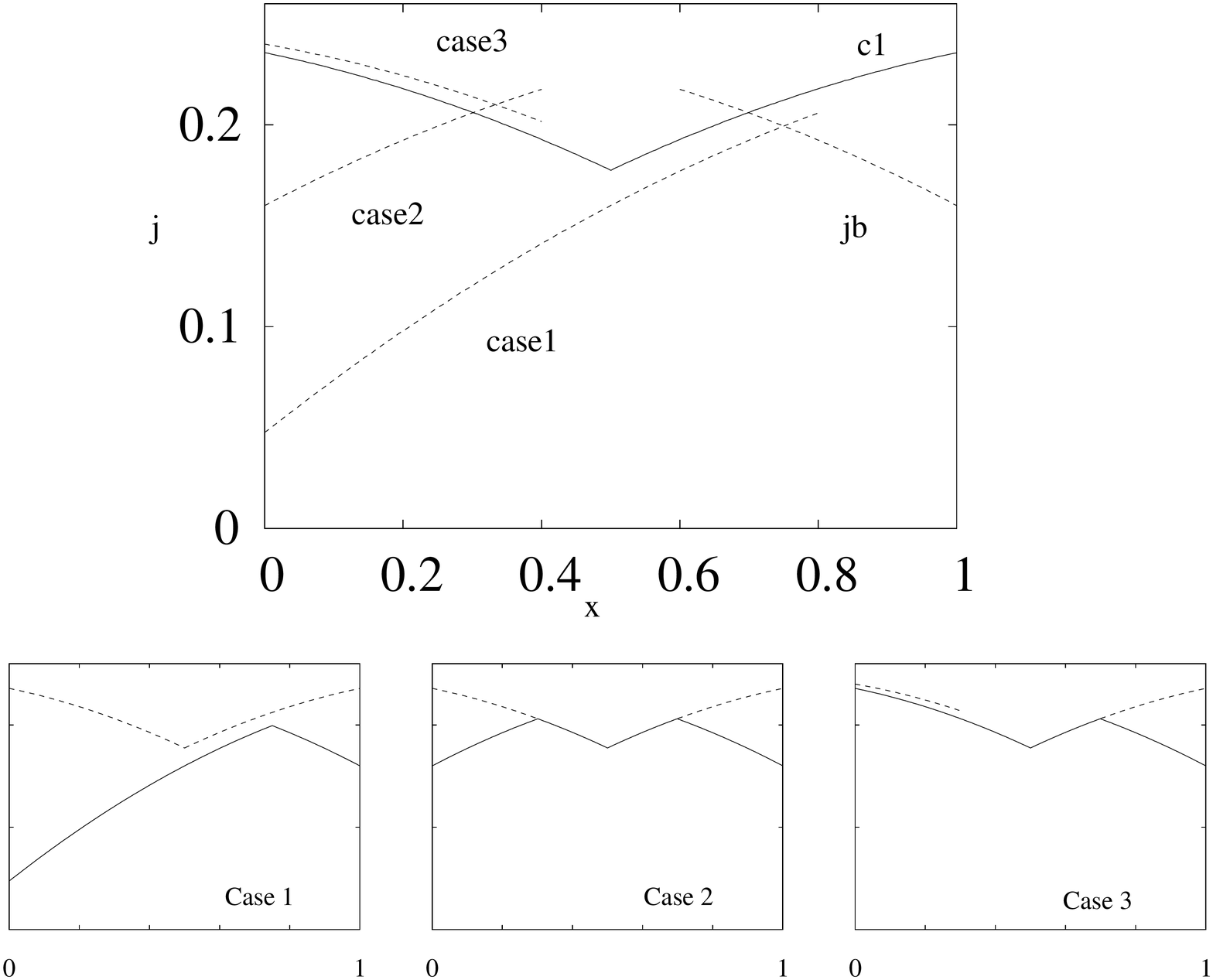}
    \caption{ \label{fig:buildcurr} Construction of the current
      profile. Top: For given carrying
      capacity ${\cal C}_1(x)$ (solid line) and fixed exit rate $\beta=0.2$, various
      scenarios arise when the entrance rate $\alpha=\{0.05,0.2,0.6\}$ is varied.
      Bottom: Three different
      profiles are shown in the small graphs (solid line: global
      current profile, dashed line: defect and boundary currents)
      emerging from three different left boundary conditions.
      Parameters are $x_{\d}=1/2$, $q=0.3$, $\Omega=0.3$,
      $\beta=0.2$, $\alpha=\{0.05,0.2,0.6\}$.}
    \end{center}
\end{figure}

Let us first consider the case where the carrying capacity is entirely
determined by the defect, i.e.\ ${\cal C}_1(x)=j_d(x)$.  In this
situation, sketched in Fig.~\ref{fig:excurrb}a, the bottleneck is
\emph{strong} enough and always imposes a current $j_d(x)<j^{*}$. Then
we may distinguish between three cases depending on the magnitude of
the current imposed by the left and right boundary (see
Fig.~\ref{fig:buildcurr}).

There are two extreme cases. The entrance and exit boundary currents
$j_\alpha$ and $j_\beta$ exceed the carrying capacity ${\cal C}_1(x)$,
the current through the system settled at the value $j_d(x)$ imposed
by the defect (this is situation for the left subsystem presented as
case 3 in Fig.~\ref{fig:buildcurr}). Thus, the density exhibits the
piecewise profile given by Eq.~(\ref{rho_d}). As the current is
independent of the left and right boundaries, this phase is termed
pure bottleneck phase (BP).  In contrast, for low entrance and exit
rates one recovers the TASEP/LK density profile perturbed by a local
spike or a dip (see previous section).  This is presented as case 1 in
Fig.~\ref{fig:buildcurr}.

In intermediate regions of the parameter space, a situation like the
one presented as case 2 in Fig.~\ref{fig:buildcurr}, appears.
Similarly to what happens for the homogeneous TASEP/LK, when the
densities $\rho_{\alpha}$, $\rho_{\beta}$ and $\rho_d$ cannot be
matched continuously, shocks form in the density profile and then we
have coexistence of several phases. This can happen either on the
left or right subsystem. The positions of the shocks follow from the
local conservation of the current, i.e.\ 
$j_{\alpha}(x_{\w}^{\l})=j_d(x_{\w}^{\l})$ (on L) and
$j_{\beta}(x_{\w}^{\r})=j_d(x_{\w}^{\r})$ (on R).
With Eqs.~(\ref{jd})-(\ref{jb}), one finds
\begin{eqnarray}
\label{xwL}
x_{\w}^{\l}&=& \frac{1}{2}\left( x_{\d} +\frac{1}{\Omega}\left\{\frac 1 2-\delta -\alpha\right\}\right)\, , \\
\label{xwR}
x_{\w}^{\r}&=& \frac{1}{2}\left( 1+x_{\d} +\frac{1}{\Omega}\left\{\beta+\delta-\frac 1 2 \right\}\right). 
\end{eqnarray}
The conditions for having domain walls within the two subsystems are
$0<x_{\w}^{\l}<x_{\d}$ and $x_{\d}<x_{\w}^{\r}<1$. This translates into the
following conditions for the entrance and exit rates:
\begin{eqnarray}
  \frac1 2-\delta-\Omega x_{\d} <& \alpha&< \frac1 2 -\delta+\Omega x_{\d}\, ,\\
  \frac1 2-\delta-\Omega(1- x_{\d}) <& \beta& < \frac1 2-\delta+\Omega(1- x_{\d})\, .
\end{eqnarray}
%
These conditions allow to identify two values of the boundary rates
$\alpha_c^-$ and $\beta_c^-$ for which a domain wall enters
respectively the left or right subsystem and two values $\alpha_c^+$
and $\beta_c^+$ for which the domain wall enters the left or right
subsystem:
\begin{eqnarray}
  \label{eq:critab}
  \alpha_c^\pm\equiv \frac 1 2-\delta\pm\Omega x_{\d}\, ,\\
  \beta_c^\pm\equiv \frac1 2-\delta\pm\Omega(1- x_{\d})\, .
\end{eqnarray} 
Within the range defined by these critical rates, a shock is
localized in each subsystem.
At values above $\alpha_c^-$ and $\beta_c^-$ the defect becomes
relevant and at least part of the density and current profile is
dictated by the defect. This suggests to term the corresponding
regions of the parameters space \emph{bottleneck phases}.

\begin{figure}[!ht]
  \psfrag{x}{$x$} \psfrag{r}[][][1][-90]{\hspace{5mm}$\rho$}
   \begin{tabular}{lr}
     \psfrag{a}{(a) LD-BP}
     \hspace{-5mm}\includegraphics[width=0.55\columnwidth]{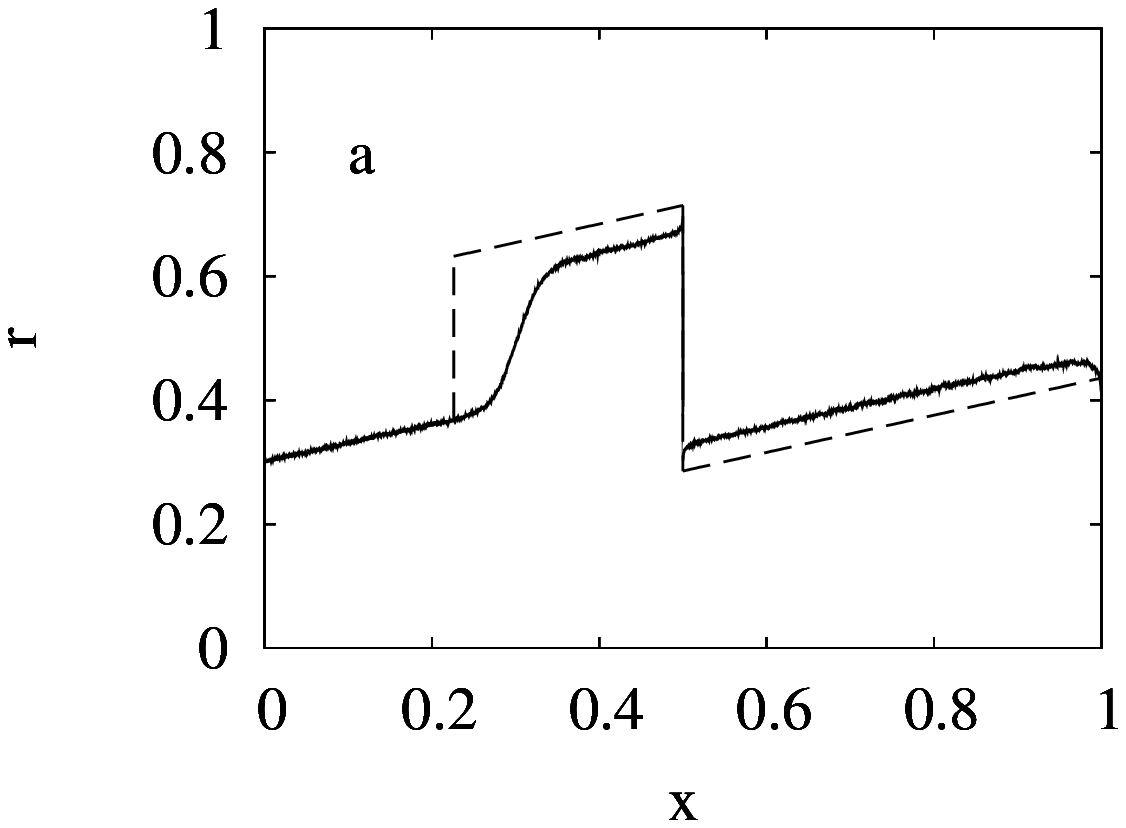}&
     \psfrag{a}{(b) BP}
     \hspace{-5mm}\includegraphics[width=0.55\columnwidth]{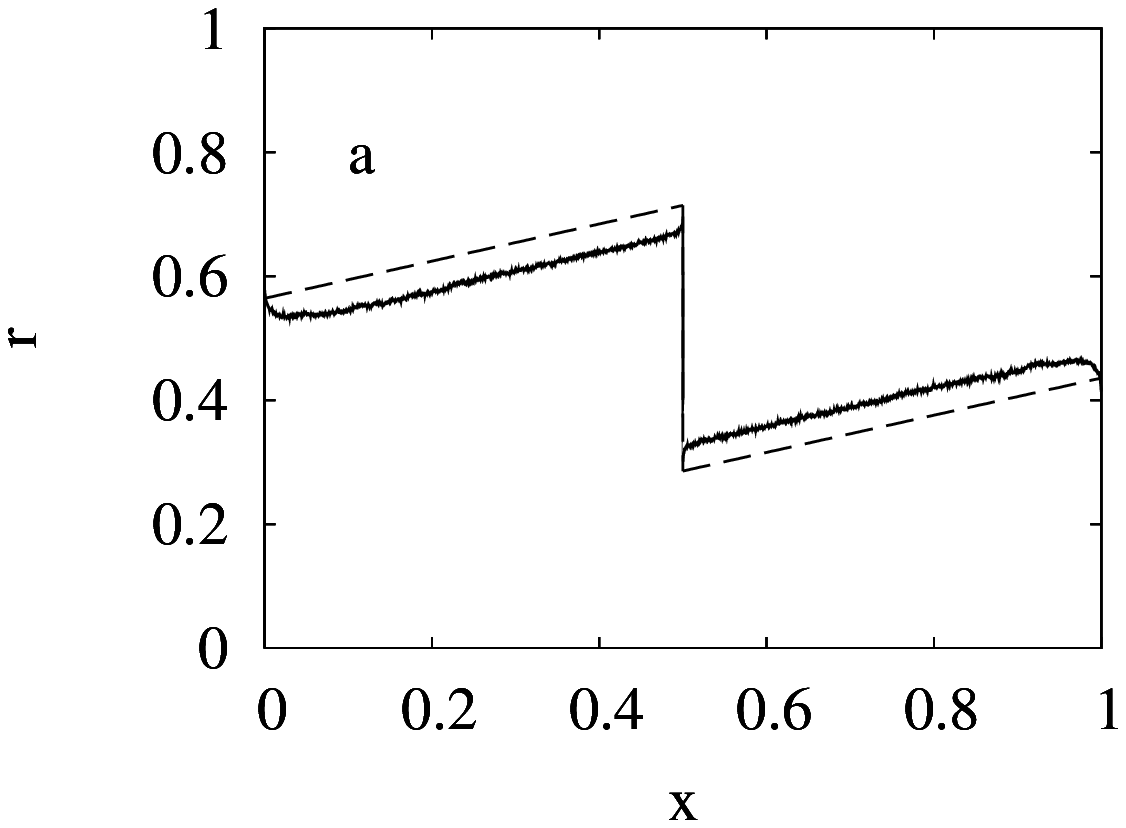}\\
     \psfrag{a}{(c) LD-BP-HD}
     \hspace{-5mm}\includegraphics[width=0.55\columnwidth]{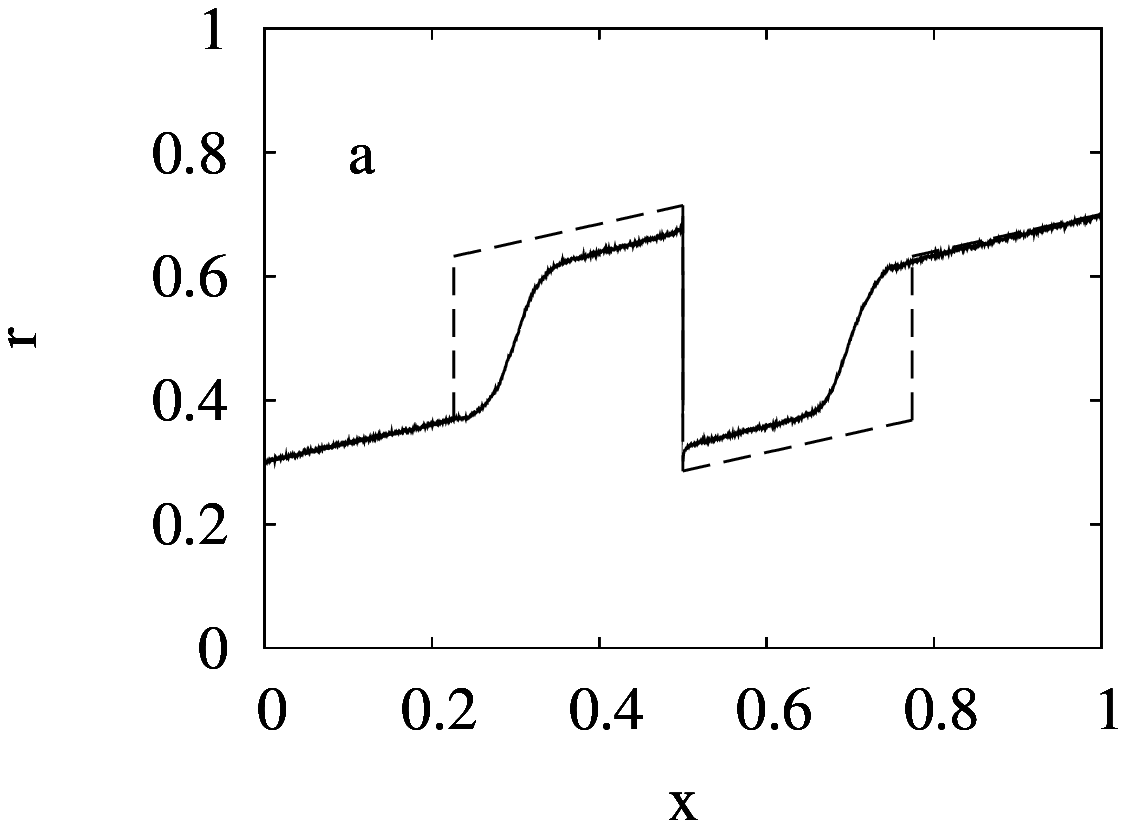}&
     \psfrag{a}{(d) BP-HD}
     \hspace{-5mm}\includegraphics[width=0.55\columnwidth]{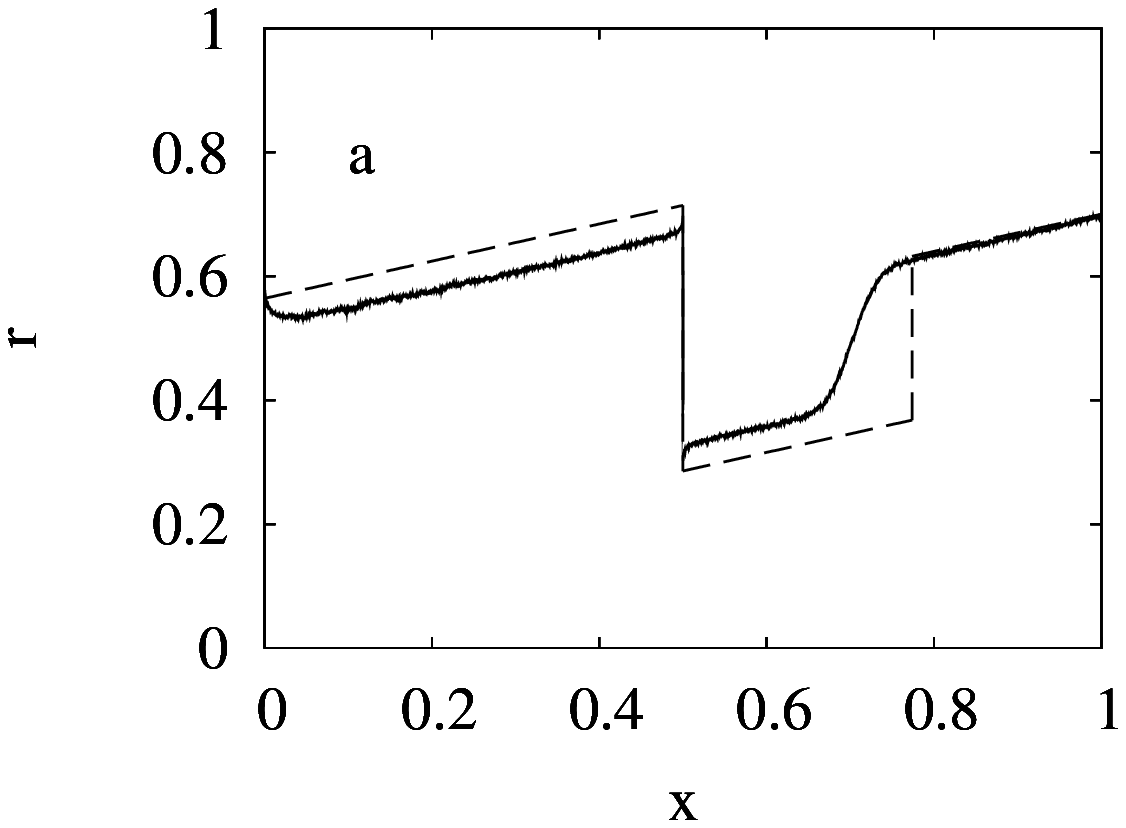}
   \end{tabular}
   \caption{\label{fig:k1eg} \quad Examples of density profiles in the
     bottleneck phases for a carrying capacity ${\cal C}(x)={\cal
       C}_1(x)$. Stochastic simulations (continuous line) are compared
     to analytical mean-field predictions (dashed line). The system
     size is $N=4096$ and the parameters are $q=0.3$, $x_{\d}=0.5$,
     $\Omega_D=0.3$, LD-BP: $(\alpha,\beta)=(0.3,0.6)$, BP:
     $(\alpha,\beta)=(0.6,0.6)$, LD-BP-HD: $(\alpha,\beta)=(0.3,0.3)$
     and BP-HD: $(\alpha,\beta)=(0.6,0.3)$.}
\end{figure}

Let us now construct the density profiles in the various bottleneck
phases focusing first on the left subsystem.  Depending on the
entrance rate, one distinguishes two cases: (i) The density profile
corresponding to  case 2 in Fig.~\ref{fig:buildcurr}, arising when 
$\alpha_c^-<\alpha<\alpha_c^+$,  is presented
in Fig.~\ref{fig:k1eg}a (numerical simulations are discussed later).
Here, $j_{\alpha}$ intersects the left
branch of ${\cal C}_1(x)$ at $x_{\w}^{\l}$, where a domain wall forms. Then
the density (of the left subsystem) reads:
\begin{eqnarray}
\label{rhoLD-BP}
\rho^{\l}(x) =
 \begin{cases} \rho_{\alpha}(x) &\text{, $0<x<x_{\w}^{\l}$}\\
 \rho_d(x) & \text{, $x_{\w}^{\l}<x< x_{\d}$}
\end{cases}\, .
\end{eqnarray}
This coexistence phase is called LD-BP since it is characterized by
the coexistence of a low density and a bottleneck phases.  (ii) The
density profile corresponding to case 3 in Fig.~\ref{fig:buildcurr},
arising for $\alpha>\alpha_c^+$, is shown in Fig.~\ref{fig:k1eg}b.
Here, $j_{\alpha}$ is always above the left branch of ${\cal C}_1(x)$
and the defect imposes an effective high density phase (called simply
BP) corresponding to the maximal current $j_{d}$ on subsystem \l. The
corresponding density profile reads $\rho^{\l}(x)=\rho_d(x)$.

One proceeds in a similar way for the right subsystem (\r).  When
$\beta_c^-<\beta <\beta_c^+$, there is coexistence between
high-density (boundary induced) and an effective low-density (defect
induced) phases, called BP-HD (see Fig.~\ref{fig:k1eg}d). This results
in a domain wall at $x_{\w}^{\r}$ and in the density
\begin{equation}
  \rho^{\r}(x) =
  \begin{cases}
    \rho_d(x) &\text{, $x_{\d}< x<x_{\w}^{\r}$}\\
    \rho_{\beta}(x) & \text{, $x_{\w}^{\r}<x<1$}
  \end{cases}\, .
\end{equation}
When $\beta> \beta_c^+$, the defect dominates and imposes an effective
low density phase, called BP where the density is
$\rho^{\r}(x)=\rho_d(x)$.

Considering the whole system, by combining the above mixed phases (on
$\r$ and $\l$) we obtain the following four bottleneck phases:
\begin{center}
\begin{tabular}{|c||c|c|}
  \hline
  \hline
  Left$\downarrow$/Right$\rightarrow$&BP-HD&BP\\
  \hline\hline
  LD-BP& LD-BP-HD &  LD-BP    \\
  \hline
  BP &BP-HD &BP\\ \hline
\end{tabular}
\end{center}

The emergence of these four defect dominated mixed phases is the most
dramatic effect of the bottleneck when
$\delta>\max\{\delta_1,\delta_2\}$.  We have checked the predictions
of our MF theory against stochastic numerical simulations (following
the Bortz-Kalos-Lebowitz scheme for kinetic Monte Carlo
\cite{bortz-kalos-lebowitz:75}) and, as shown in Fig.~\ref{fig:k1eg},
have found good agreement (both qualitative and quantitative) with the
predictions of the MF theory. Of course, due to finite-size effects,
boundary layers form in the vicinity of the shocks and `soften' the
transition in the density profile. When the system-size is increased,
the accuracy of the MF theory improves and the density profile
displays a sharp jump.  In addition to boundary layers, as already
noted in Ref.~\cite{janowsky-lebowitz:92} for the simple TASEP
perturbed by one defect, in the bottleneck phases, one remarks a
slight but systematic deviation from the MF predictions (see e.g.
Figs.~\ref{fig:k1eg}b and \ref{fig:k1eg}d).  This effect is due to
correlations not taken into account by the mean-field approximation
and is long-range, scaling as the inverse of the distance to the
defect \cite{janowsky-lebowitz:92}.
\begin{figure}[htbp]
  \begin{center}
    \psfrag{alpha}{$\alpha$}
    \psfrag{beta}{$\beta$}
    \psfrag{a1}{$\alpha_c^-$}
    \psfrag{a2}{$\alpha_c^+$}
    \psfrag{b1}{$\beta_c^-$}
    \psfrag{b2}{$\beta_c^+$}
    \vspace{1.5cm}
    \includegraphics[width=0.8\columnwidth]{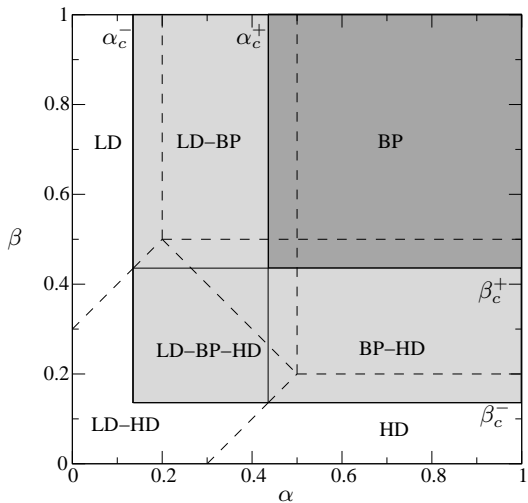}
    \caption{ \label{fig:pddef4} Phase diagram for $\Omega=0.3$,
      $q=0.4$ and $x_{\d}=1/2$, i.e.\ for the carrying capacity ${\cal
        C}_1(x)$ (large screening length).  Continuous lines are the
      phase boundaries introduced by the defect; dashed lines are the
      phase boundaries already present in the model without
      bottleneck. The shadowed region indicates the \emph{bottleneck
        phases} where the defect is relevant, the darkest one
      highlights the pure bottleneck phase (for the meaning of the
      different phases see text).}
    \end{center}
\end{figure}

The various phases are summarized in the $\alpha/\beta$-phase-diagram
of Fig.~\ref{fig:pddef4}.  In this figure, because $x_{\d}=1/2$ and
$\Omega_A=\Omega_D=\Omega$, the particle-hole symmetry (which still
holds in the presence of the bottleneck) results in the invariance of
the phase-diagram w.\ r.\ t.\ the line $\alpha=\beta$.

As can be seen from the phase-diagram of Fig.~\ref{fig:pddef4}, the
\emph{bottleneck-induced} mixed phases (LD-BP-HD, LD-BP, BP-HD and BP)
occupy the upper right part of the diagram (shadowed region in
Fig.~\ref{fig:pddef4}). Only at the borders of the phase-diagram,
corresponding to particularly low/high entrance/exit rates, one
recovers the same phases as in the defect-free TASEP/LK model (there
the defect is irrelevant). For $\alpha>\alpha_c^+$ and
$\beta>\beta_c^+$, i.e.\ in the right top corner of the phase-diagram
(darkest shadowed region in Fig.~\ref{fig:pddef4}), the entire system
is in a \emph{pure bottleneck phase}. By tuning the strength of the
defect $\delta$ and the on-off parameter $\Omega$, the phase
boundaries can be shifted to recover the short screening length case
(discussed in the following) and eventually the usual TASEP/LK
behavior (when the defect is irrelevant).  On the other hand, by
increasing the strength of the defect (or by reducing the on-off rate)
the phase boundaries move toward the axes of the phase-diagram and can
even merge with them. Actually, this occurs for a defect strength
$\tilde \delta$. Above this threshold, the whole
$\alpha/\beta$-phase-diagram is characterized by bottleneck phases
(the homogeneous LD, HD and LD-HD phases are squeezed out of the
diagram).  The critical strength $\tilde \delta$ is therefore the
maximum between the value determined by the conditions $\alpha_c^-=0$
and $\beta_c^-=0$, which leads to
\begin{equation}
  \tilde\delta=\max\left\{\frac{1}{2}-\delta_1,\,  \frac{1}{2}-\delta_2\right\}\, .
\end{equation}
Note that for a too strong defect, $\delta>\tilde\delta$, the control
of the system from the boundaries is lost.

\subsection{Short screening length}
\label{sec:flatcarrying}
We now consider the case where the screening length is short so that
the carrying capacity reaches the value $j^*=1/4$ in the bulk of the
system.

\subsubsection{Symmetric screening}
We first consider the case illustrated in Fig.~\ref{fig:excurrb}b, the
carrying capacity loses the trace of the defect in the system, at a
distance larger than the screening length:
\begin{equation}
\label{eq:c2}
 {\cal C}_2(x)=\begin{cases} j_d(x)&, |x-x_{\d}|<\xi\\
   \frac{1}{4} &, \textrm{else}
\end{cases}\, .
\end{equation}
This situation arises when $\delta<\min\{\delta_1,\delta_2\}$ (with
fixed $\Omega$) or, equivalently, when
$\Omega>\max\{\Omega_1,\Omega_2\}$ (with fixed $\delta$).  This
carrying capacity corresponds to the richest case in terms of new
bottleneck phases, due to the profile of ${\cal C}_2$ characterized by
four distinct regions.
 
As in the previous situation, the bottleneck is relevant and induces
new mixed phases when $\alpha>\alpha_c^{-}$ and $\beta>\beta_c^{-}$,
where these critical values are again given by
Eqs.~(\ref{eq:critab}). Elsewhere in the parameter space, the
homogeneous TASEP/LK profiles locally perturbed by a spike (or a
dip) are recovered. 

\begin{figure}
 \begin{center}
   \psfrag{alpha}{$\alpha$}
   \psfrag{beta}{$\beta$}
   \psfrag{a1}{$\alpha_c^{-}$}
   \psfrag{a2}{$\alpha_c'$}
   \psfrag{a*}{$\alpha^*$}
   \psfrag{b1}{$\beta_c^{-}$}
   \psfrag{b2}{$\beta_c'$}
   \psfrag{b*}{$\beta^*$}
   \vspace{15mm}
   \includegraphics[width=0.8\columnwidth]{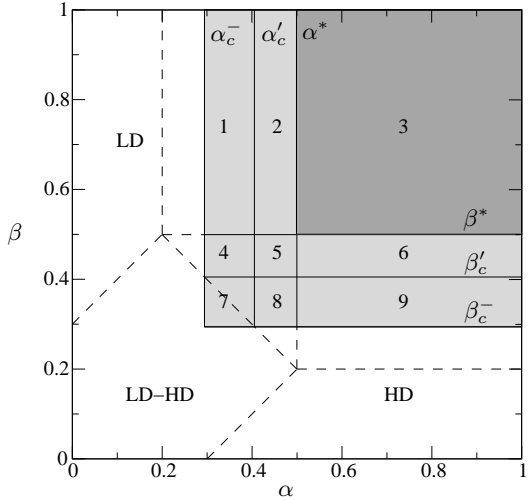}
   \caption{ \label{fig:pddef3}Phase-diagram for
     $\Omega=0.3$, $x_{\d}=1/2$ and $q=0.8$, i.e.\ for the carrying
     capacity ${\cal C}_2(x)$ (short screening length).  Continuous
     lines are the phase boundaries introduced by the defect; dashed
     lines are the phase boundaries already present in the model
     without bottleneck.  The shadowed region indicates the
     \emph{bottleneck phases} where the defect is relevant, the
     darkest one highlights the bottleneck phase independent of the
     entrance ($\alpha$) and exit ($\beta$) rates (see the text and
     Tab.~\ref{tab:c1}).}
   \end{center}
\end{figure}

When the system is driven above its carrying capacity
($\alpha,\beta>1/2$), it exhibits the current profile given in
Eq.~(\ref{eq:c2}), corresponding to a density profile
\begin{eqnarray}
  \rho^{\l}(x)=
  \begin{cases}
    \rho^{*} &, x_{\d}-x>\xi \\
    \rho_{d}(x)&, x_{\d}-x<\xi \end{cases}.
\end{eqnarray}
As the current profile of Eq.~(\ref{eq:c2}) reaches the MC value,
contrary to the case of a large screening length, this is no longer a
pure bottleneck phase but corresponds to a MC-BP-MC phase (see dark
shadowed region in Fig.~\ref{fig:pddef3}).

Between these two extremal situations, more intricate mixed phases
appear. As in the previous case, phases characterized by shocks appear
when the boundary currents match $j_d$.

Contrary to the previous situation, where $j_d$ covered the whole
subsystem, now the defect imposed current extends up to a distance
equal to the screening length from the bottleneck. Hence, shocks can
emerge only in a macroscopic region, $x_{\d}-\xi<x_{\w}^{\l}<x_{\d}$ and 
$x_{\d}<x_{\w}^{\r}<x_{\d}+\xi$, in the vicinity of the defect.  These
conditions translate into new critical values of the entrance and exit
rates, namely
\begin{eqnarray}
\label{prime}
\alpha_c'&=& \frac{1}{1+q}-\Omega x_{\d}\, ,\quad \beta_c'=\frac{1}{1+q}-\Omega (1-x_{\d})
\label{xr}
\end{eqnarray}
%

\begin{figure}
 \psfrag{j*}{$j^*$} \psfrag{ja}{$j_{\alpha}$}
 \psfrag{jb}{$j_{\beta}$} \psfrag{jl}{$j_{\beta_{\rm eff}}$}
 \psfrag{jr}{$j_{\alpha_{\rm eff}}$} \psfrag{r}[][][1][-90]{$\rho$} \psfrag{x}{$x$}
 \psfrag{j}[][][1][-90]{$j$} \psfrag{xd}[t][][0.9]{$x_{\d}$}
 \psfrag{x1}[][][0.9]{$\xi$} \psfrag{x2}[][][0.9]{$\xi$}
 \psfrag{xl}[][][0.9]{$x_I$} \psfrag{xr}[][][0.9]{$x_{II}$}
 \psfrag{jl}{$j_{\beta_{\rm eff}}$} \psfrag{jr}{$j_{\alpha_{\rm
       eff}}$}
  \includegraphics[width=0.8\columnwidth]{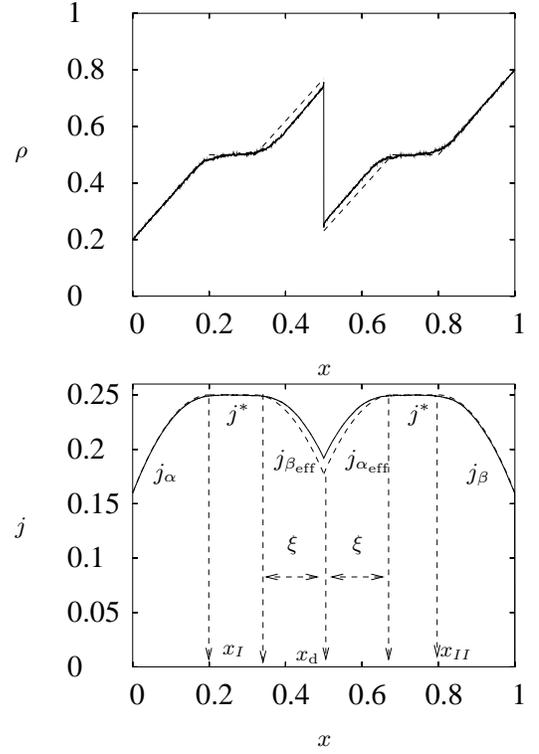}
   \caption{\label{fig:eg2} \quad Examples of density profile (top)
     and current (bottom) for the system in the LD-MC-BP-MC-HD phase:
     the defect is relevant and the carrying capacity is ${\cal
       C}_2(x)$.  stochastic simulations (continuous line) are
     compared to analytical mean-field predictions (dashed line). The
     system size is $N=4096$, $K=1$, $\Omega=\Omega_D=1.5$, $q=0.3$,
     $(\alpha,\beta)=(0.2.0.2)$ and $x_{\d}=1/2$: one can clearly
     distinguish the various phases and note a discontinuity in the
     density profile in the proximity of the defect, while the current
     profile is continuous.  }
\end{figure}

As a new scenario, here the boundary currents can reach the upper-bound
$j^*=1/4$ on the subsystems L and R when
$j_{\alpha}(x^{\l})=j_{\beta}(x^{\r})=1/4$. With Eqs.~(\ref{ja}) and
(\ref{jb}), one finds that this occurs at
\begin{eqnarray}
\label{xlxr}
x^{\l}= \frac{1-2\alpha}{2\Omega}\, ,\quad x^{\r}=1 - \frac{1-2\beta}{2\Omega}\, .
\end{eqnarray}
We note that these quantities are independent of the properties of the defect, 
which is thus screened.
Let us consider a density profile which exemplifies the richness of
this case. When $\alpha_c'<\alpha<\alpha^*=1/2$ the resulting current
$j_\alpha(x)$ saturates at the value $1/4$, while the density reads
$\rho^{\l}(x)=1/2$ for $x^{\l}<x<x_{\d}-\xi$.  Similarly, on the
subsystem $\r$, when $\beta_c'<\beta<\beta^*=1/2$ and the current
$j_\beta(x)$ saturates at the value $1/4$, the density is
$\rho^{\l}(x)=1/2$ for $x_{\d}+\xi<x<x^{\r}$.  Within the range of the
screening length, $x_{\d}-\xi<x<x_{\d}+\xi$, the carrying capacity and the
current flowing through the system coincide with $j_d(x)$ and the
density is given by Eq.~(\ref{rho_d}). Summarizing, in this case the
density profile is piecewise and one distinguishes five regions:
\begin{eqnarray}
\label{6P}
\rho(x)=\begin{cases}
  \rho_{\alpha}(x) &,  0<x<x^{\l} \\
  \rho^*&,  x^{\l}< x<x_{\d}-\xi \\
  \rho_{d}(x) &,  x_{\d}-\xi <x<x_{\d}+\xi \\
  \rho^* &, x_{\d}+\xi< x<x^{\r} \\
  \rho_{\beta}(x) &,  x^{\r}<x<1
\end{cases}
\end{eqnarray}
This case, denoted LD-MC-BP-MC-HD, is illustrated in
Fig.~\ref{fig:eg2} and corresponds to the coexistence of a low-density
and high-density (on L and R, respectively), two maximal current (one
in both subsystems) and a bottleneck-induced mixed phases.

The three possible scenarios for the bottleneck phases on the two
sub-lattices results in nine mixed phases on the whole
system, as summarized in Tab.~\ref{tab:c1}.
\begin{table*}
 \begin{tabular}{|c||c|c|c|}
   \hline
   Left$\downarrow$/Right$\rightarrow$&BP-HD&BP-MC-HD&BP-MC\\
   \hline\hline
   LD-BP& LD-BP-HD$^7$ &  LD-BP-MC-HD$^4$ &LD-BP-MC$^1$    \\
   \hline
   LD-MC-BP &LD-MC-BP-HD$^8$ &LD-MC-BP-MC-HD$^5$ &LD-MC-BP-MC$^2$   \\
   \hline
   MC-BP &  MC-BP-HD$^9$ &MC-BP-MC-HD$^6$ &MC-BP-MC$^3$ \\  \hline
 \end{tabular}
 \caption{\label{tab:c1}Bottleneck phases for the case ${\cal C}(x)={\cal C}_2(x)$. For each phase, the label (1-9) refers to a given region of the (shadowed part) of the phase-diagram represented in Fig.~\ref{fig:pddef3}.}
\end{table*}

We have also checked our MF predictions against stochastic numerical
simulations, as illustrated in Fig.~\ref{fig:eg2}.  Again, we have
found qualitative and quantitative agreement. The small deviations
from the MF theory observed in Fig.~\ref{fig:eg2} can be explained
along the above discussion on the role of the correlations and
finite-size effects.

The $\alpha/\beta$-phase-diagram corresponding to a system with a
carrying capacity ${\cal C}_2(x)$ is shown in Fig.~\ref{fig:pddef3}
and characterized by the transitions lines corresponding to the
critical values $\alpha_c^-, \alpha_c', \alpha^*$ and $\beta_c^-,
\beta_c', \beta^*$.  The nine bottleneck-induced phases appear in the
core of the phase diagram (shadowed region in Fig.~\ref{fig:pddef3}).
Contrary to the phase-diagram obtained with ${\cal C}={\cal C}_1$,
instead of a pure bottleneck phase, here one finds a MC-BP-MC
coexistence phase, which is independent of the entrance and exit.
boundaries (right top corner of Fig.~\ref{fig:pddef3}). This phase,
obtained when $\alpha>\alpha^*$ and $\beta>\beta^*$, is the
`bottleneck analogous' of the MC phase in the homogeneous TASEP/LK
system.

\subsubsection{Asymmetric (partial) screening }

To conclude the study of the case $\Omega_A=\Omega_D$, let us briefly
consider the scenarios where the system displays a carrying capacity
of type ${\cal C}_3 (x)$ or ${\cal C}_4 (x)$. This is possible when
the defect is not at the center of the system. In fact, the carrying
capacity of the system is ${\cal C}_3 (x)$ when
$\delta_1<\delta<\delta_2$ and is ${\cal C}_4 (x)$ when
$\delta_2<\delta<\delta_1$.  As one can infer from the profile of
${\cal C}(x)={\cal C}_3 (x)$ [see Fig.~ c], on the subsystem L (R)
this case is identical to that discussed for ${\cal C}(x)={\cal
  C}_1(x)$ (${\cal C}(x)={\cal C}_2(x)$). Therefore, it directly
follows from the above tables that there are six new phases for the
system displaying a carrying capacity ${\cal C}_3 (x)$.  The latter
are summarized in the following table:
%
\begin{center}
\begin{tabular}{|c||c|c|c|}
 \hline
 \hline
 Left$\downarrow$/Right$\rightarrow$&BP-HD&BP-MC-HD&BP-MC\\
 \hline\hline
 LD-BP& LD-BP-HD &  LD-BP-MC-HD &LD-BP-MC    \\
 \hline
 BP &  BP-HD & BP-MC-HD & BP-MC \\  \hline
\end{tabular}
\end{center}
%

Similarly, six bottleneck phases are also obtained for a carrying
capacity ${\cal C}(x)={\cal C}_4 (x)$, as it is clear from the table
below:

\begin{center}
\begin{tabular}{|c||c|c|c|}
 \hline
 \hline
 Left$\downarrow$/Right$\rightarrow$&BP-HD&BP\\
 \hline\hline
 LD-BP& LD-BP-HD &  LD-BP     \\
 \hline
 LD-MC-BP &LD-MC-BP-HD &LD-MC-BP    \\
 \hline
 MC-BP &  MC-BP-HD &MC-BP\\  \hline
\end{tabular}
\end{center}
The corresponding  phase-diagrams  directly follow from those discussed above for the
carrying capacities ${\cal C}_1 (x)$ and ${\cal C}_2 (x)$. 

\subsection{Topological features}
\begin{figure}[htbp]
  \begin{center} \psfrag{a}[][][1.2]{$\alpha$} \psfrag{q}{$q$}
    \psfrag{q*}{$\delta_1$} \psfrag{Q}{${\cal Q}$} \psfrag{P1}{${\cal P}$}
    \psfrag{a1}{$\alpha_c^-$} \psfrag{a2}{$\alpha_c^+$}
    \psfrag{a2p}{$\alpha_c'$} \psfrag{a*}{$\alpha^*$}
    \psfrag{at}{$\frac 1 2 -\Omega_D$} \psfrag{ 0}{$0$} \psfrag{
      0.2}{$0.2$} \psfrag{ 0.4}{$0.4$} \psfrag{ 0.6}{$0.6$} \psfrag{
      0.8}{$0.8$} \psfrag{ 1}{$1$} \psfrag{ 0.5}{$0.5$}
    \psfrag{qt}{$\tilde \delta$} \psfrag{d}[][][1.2]{$\delta$}
    \vspace{5mm}
  \includegraphics[width=1\columnwidth]{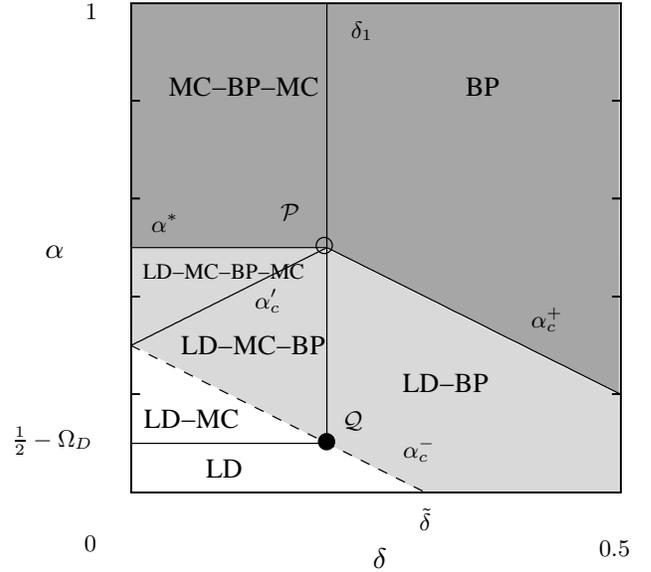}
  \caption{ \label{fig:defq} Cut of the phase-diagram in $\delta$ and
    $\alpha$ (parameters: $\Omega=0.4$, $x_{\d}=1/2$, $\beta>1/2$).
    Solid lines identify usual phase transitions, while the dashed
    line the peculiar transition between relevant and irrelevant
    defect. The multiple coexistence points ${\cal P}$ and ${\cal Q}$
    are shown in the graph (see main text).}
\end{center}
\end{figure}

The above discussion has shown that the sole presence of a localized
bottleneck is responsible for the emergence of new (sub-)phases and
drastic topological changes of the $\alpha/\beta$-phase-diagram. Here,
we aim to discuss further the important structural changes induced by
the bottleneck in the density and current profiles of the TASEP/LK by
considering the $\alpha/\delta$-phase-diagram (see
Fig.~\ref{fig:defq}).

We have already seen that the most appealing properties of the
phase-diagrams of the model under consideration are the new bottleneck
(sub)phases.  As summarized in Figs.~\ref{fig:pddef4}
and~\ref{fig:pddef3} a large portion of the
$\alpha/\beta$-phase-diagrams is dominated by the defect properties.
The phase boundaries separating the usual phases from the bottleneck
(i.e.\ $\alpha=\alpha_c^-$ and $\beta=\beta_c^-$) are straight lines
since they depend only on the detachment rate and on the strength and
position of the defect, but not on the entrance exit rates.  

In the simple TASEP, the transitions between the usual phases and the
bottleneck phases can be considered as discontinuous, since the
influence of the defect changes abruptly from a local peak to a global
step (see Fig.~\ref{fig:spike1}).  In the case of TASEP/LK with
bottleneck, a shock enters continuously the system and the transition
can be consider indeed continuous, as it was the in the simple
TASEP/LK (dashed line in Fig.~\ref{fig:defq}).

The phase boundaries $\alpha_c^+$ ($\beta_c^+$), for $\delta>\delta_1$
($\delta>q_2$), and $\alpha^*=1/2$ ($\beta^*=1/2$), for
$\delta<\delta_1$ ($\delta<\delta_2$), separate the point where the
density profile is dominated by the defect from the one where the
boundary currents still play a role. Considering the position of the
domain walls $x_{\w}^{\l}$ and $x_{\w}^{\r}$ as the order parameters,
the transitions are all continuous. A similar transition is also
identified by the boundary $\alpha_c'$ ($\beta_c'$) since the matching
point (no longer a shock) $x^{\l}$ ($x^{\r}$) enters the system
continuously.

The physical modifications in the system following the transition
between a carrying capacity ${\cal C}_1(x)$ and ${\cal C}_2(x)$ can be
observed if we plot the cut of the phase-diagram in $\delta$ and
$\alpha$ (choosing $\beta>1/2$), as done in Fig.~\ref{fig:defq}.  In
this graph, the line $\alpha_c^-$ identifies the transition between
relevant an irrelevant defect, while $\alpha_c^+$, $\alpha_c'$ and
$\alpha^*=1/2$ the transitions between bottleneck phases.  The line
$\alpha=1/2-\Omega_D$ was present in the pure TASEP/LK. The line
$\delta=\delta_1$ identifies the continuous transition between the two
different carrying capacities, ${\cal C}_1(x)$ and ${\cal C}_2(x)$.
There are two relevant points in the phase diagram where different
phases coexist, indicated with ${\cal P}$ (five phases) and ${\cal Q}$
(four phases) in Fig.~\ref{fig:defq}.

\subsection{Results and phase-diagrams when $\Omega_A\neq\Omega_D$}
\label{sec:result2}
\begin{figure}[htbp]
  \begin{center}
    \psfrag{alpha}{$\alpha$}
    \psfrag{beta}{$\beta$}
    \psfrag{a1}{}
    \psfrag{a2}{}
    \psfrag{b1}{}
    \psfrag{b2}{}
    \vspace{10mm}
    \includegraphics[width=0.8\columnwidth]{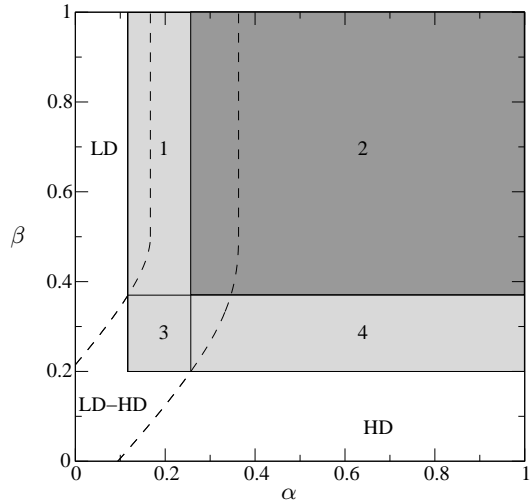}
    \caption{ \label{fig:pddef1} Phase diagram for $K=2$,
      $\Omega_D=0.1$, $x_{\d}=1/2$ and $q=0.3$, i.e.\ for the ${\cal
        C}_1(x)$-like carrying capacity.  Continuous lines are the
      phase boundaries introduced by the defect (BP); dashed lines are
      the phase boundaries already present in the model without
      bottleneck.  The shadowed region indicates the \emph{bottleneck
        phases} where the defect is relevant, the darkest one
      highlights the pure bottleneck phase.  The numbers stem for the
      different phases: LD-BP (1), BP (2), LD-BP-HD (3) and BP-HD (4).
      For the description of the phases see the text and
      Figs.~\ref{fig:eg3}.}
  \end{center}
\end{figure}

\begin{figure}
  \begin{center}
    \psfrag{x}{$x$} \psfrag{r}[][][1][-90]{\hspace{5mm}$\rho$}
    \psfrag{j}[][][1][-90]{\hspace{5mm}$j$}
    \vspace{1cm}
    \begin{tabular}{lr}
      \hspace{-5mm}\includegraphics[width=0.55\columnwidth]{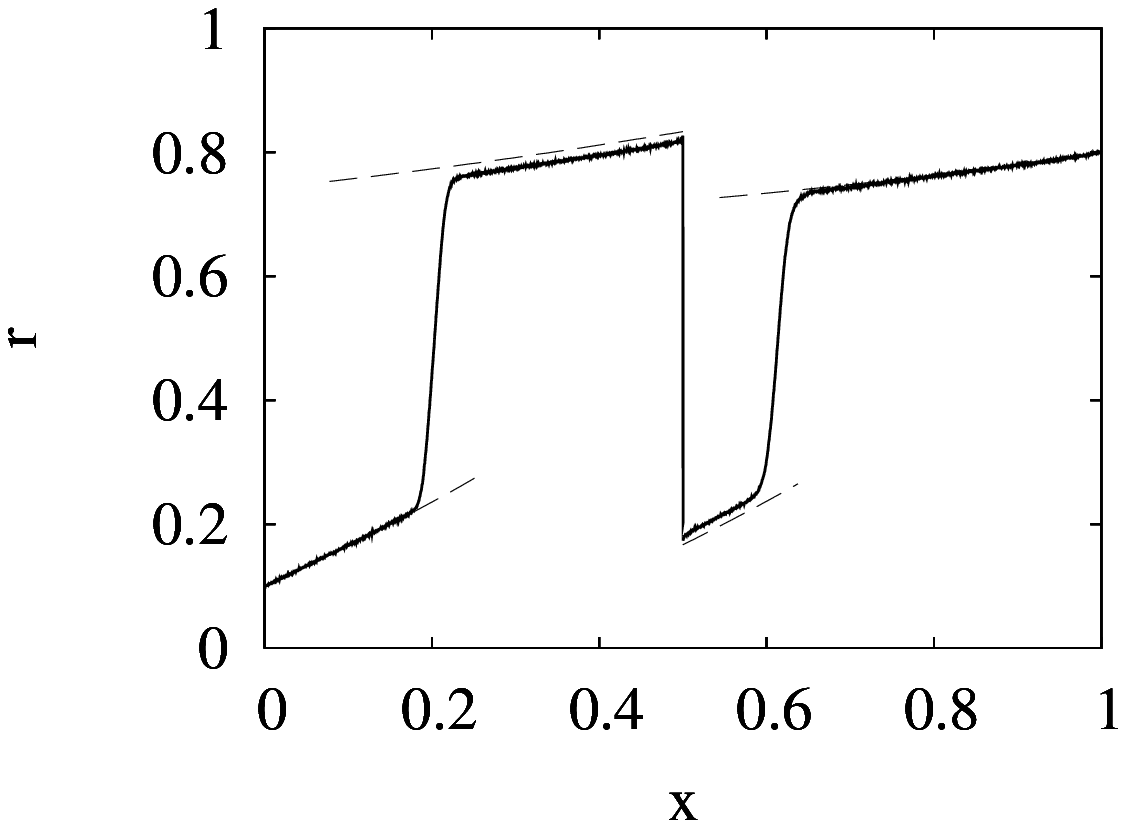}&
      \hspace{-5mm}\includegraphics[width=0.55\columnwidth]{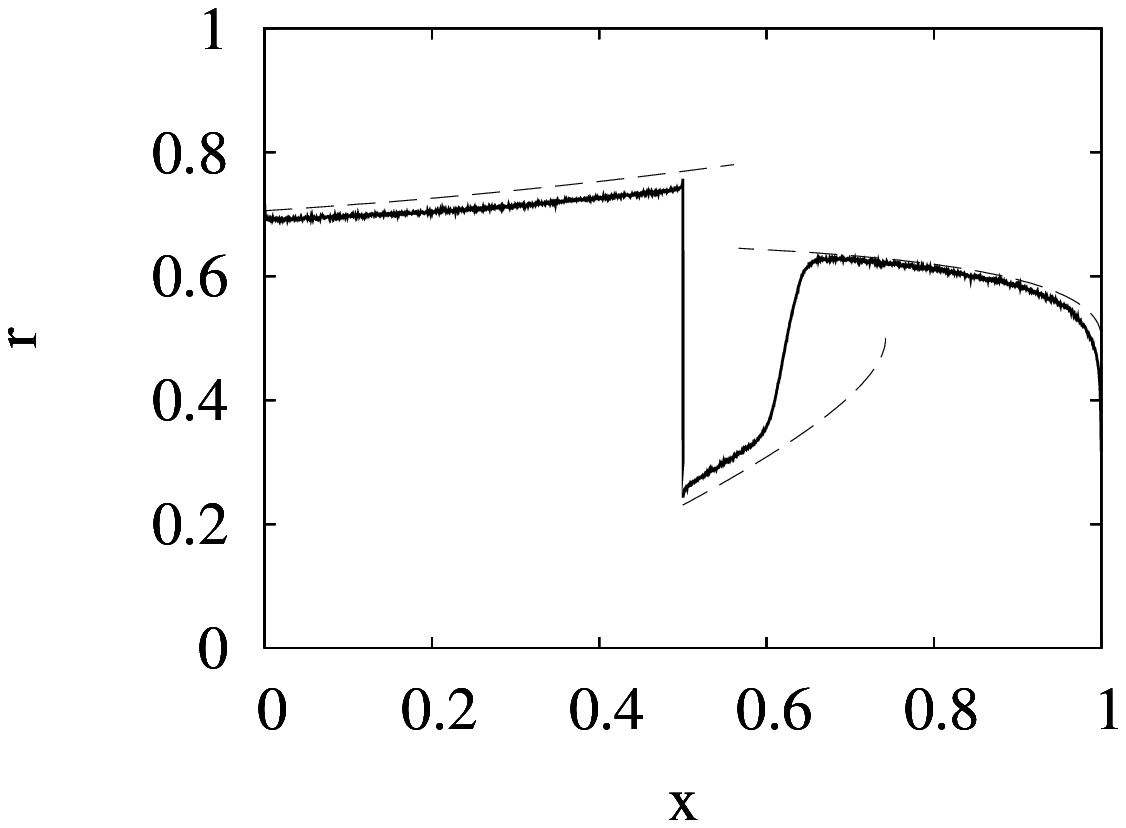}\\
      \hspace{-5mm}\includegraphics[width=0.55\columnwidth]{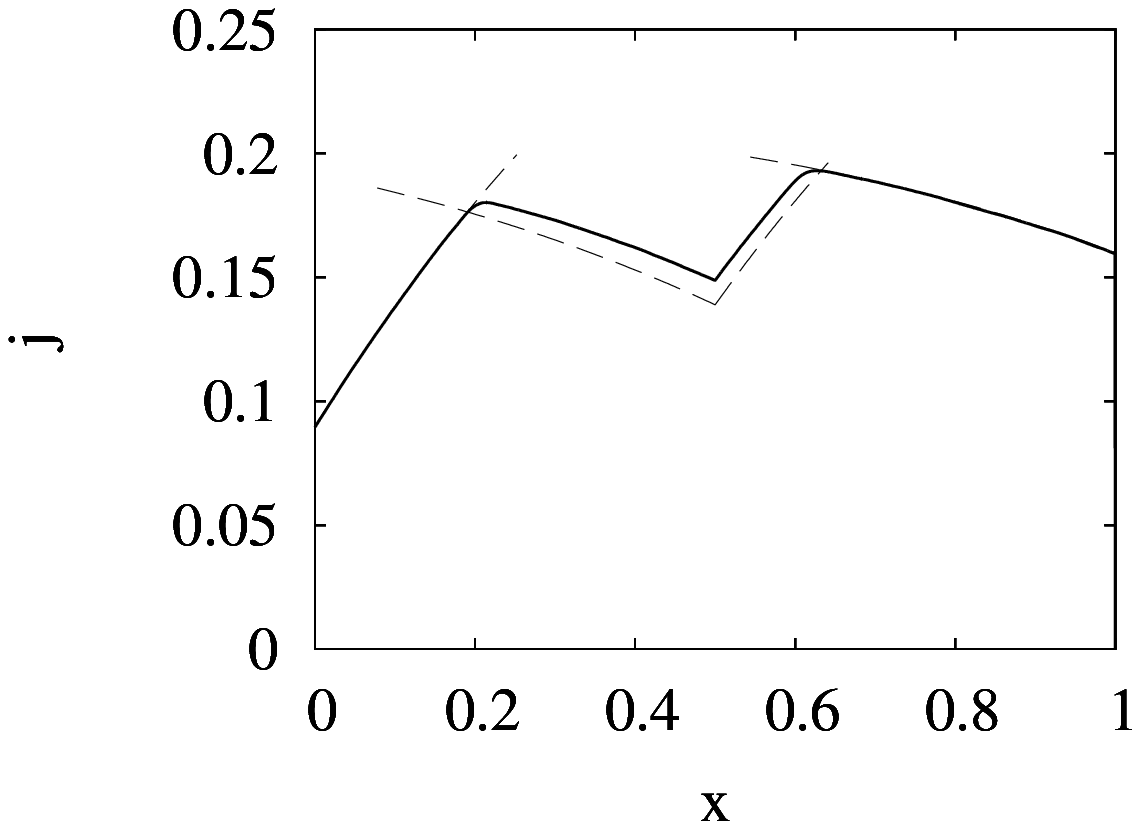}&
      \hspace{-5mm}\includegraphics[width=0.55\columnwidth]{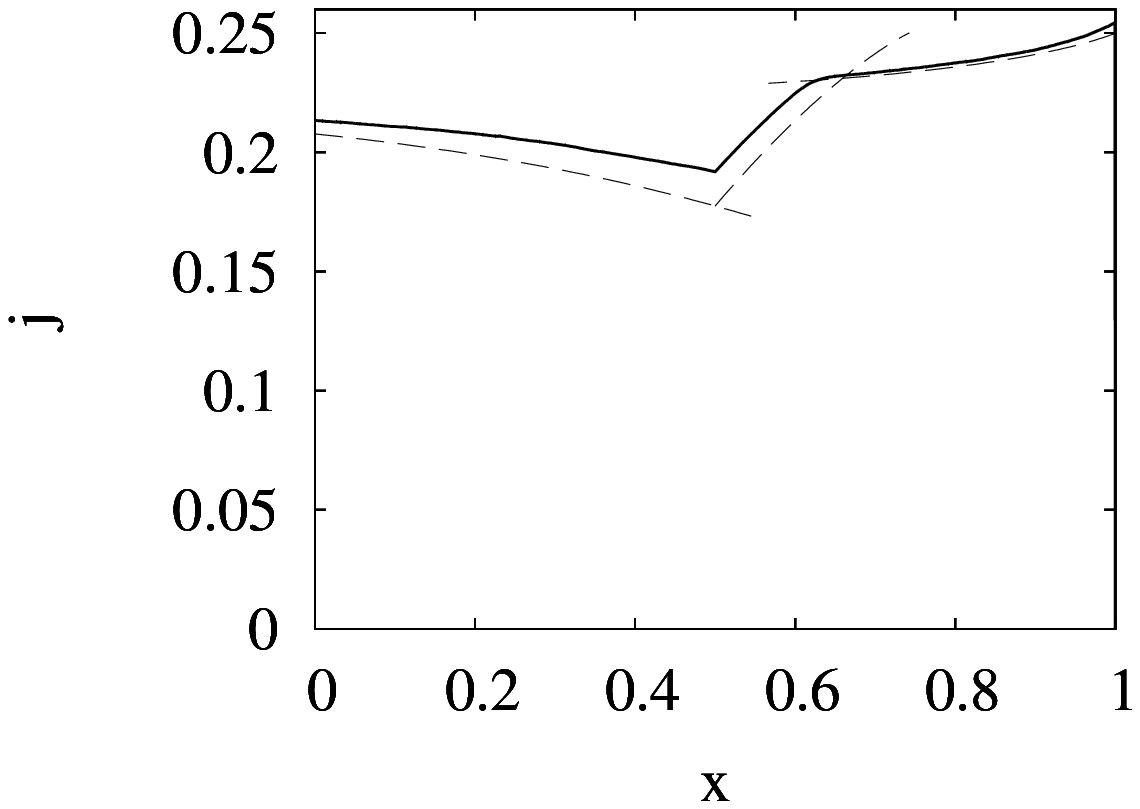}
    \end{tabular}
    \caption{\label{fig:eg3} \quad Two examples of density profile and
      current in the LD-BP-HD for the ${\cal C}_1(x)$-like carrying
      density capacity (left) and BP-MC phase for the ${\cal C}_3(x)$
      (right). Stochastic simulations (continuous line) are compared
      to analytical mean-field predictions (dashed line).  The system
      size is $N=4096$ and the parameters are $q=0.2$, $x_{\d}=1/2$,
      $K=2$, $\Omega_D=0.3$, $(\alpha,\beta)=(0.1.0.2)$ (left);
      $q=0.3$, $x_{\d}=1/2$, $K=2$, $\Omega_D=0.3$,
      $(\alpha,\beta)=(0.8,0.8)$ (right).}
      \end{center}
\end{figure}

In this Subsection we shortly discuss the case where the attachment
and detachment rates differ, $\Omega_A\neq\Omega_D$. This is
mathematically more tedious, as for the homogeneous TASEP/LK model, at
MF level, one needs to solve
Eqs.~(\ref{eq:TASEP_LK_MF},\ref{eq:TASEP_LK_BC}) which solution
implies multivalued (Lambert) functions with two real branches
\cite{parmeggiani-franosch-frey:04}. Introducing the binding constant
$K\equiv \Omega_A/\Omega_D\neq 1$, the Langmuir isotherm $\rho_I$
reads $\rho_I\equiv K/(K+1)$.  Here, having split the problem in two
subsystems one has to consider the equation
\begin{eqnarray}
  (2\rho^{\l,\r}-1)\partial_x \rho^{\l,\r}(x)=(K+1)\Omega_D[\rho_I-\rho^{\l,\r}(x)]\, ,
\end{eqnarray}
again supplemented with the (subsystems) boundary conditions
Eqs.~(\ref{eq:TASEP_LK_BC}) and (\ref{eq:effrate})
\begin{eqnarray}
  \rho^{\l}(0)=\alpha; \quad \rho^{\l}(x_{\d})=\frac{1}{1+q}\\ 
  \rho^{\r}(x_{\d})=\frac{q}{q+1}; \quad \rho^{\r}(1)=1-\beta\, .
\end{eqnarray}
We refer the readers to Ref.~\cite{parmeggiani-franosch-frey:04} for a
detailed mathematical treatment of this kind of equations, and report
our results for the phase-diagram and density profiles of the TASEP/LK
model in the presence of a bottleneck when $\Omega_D\neq \Omega_A$
(i.e.\ $K\neq 1$). As for the homogeneous model, one can take
advantage of the underlying particle-hole symmetry to restrict the
discussion to the case $K>1$ \cite{parmeggiani-franosch-frey:04}.
Except for the mathematical treatment of the MF bulk equation, we
follow the same lines as in the case $K=1$. Again, one has to
distinguish the case where the carrying capacity coincides with the
current imposed by the defect (${\cal C}(x)=j_d(x)$) from the
situation where ${\cal C}(x)$ reaches the maximal current value
$j_K^*$.  While for $K=1$ the maximal current available in the bulk is
$j^*=1/4$, here $j^*_K(x)\leq j^*$ is a non-constant space-dependent
quantity.  When ${\cal C}(x)=j_d(x)$, except some topological
asymmetries, one essentially recovers the same phase-diagram as in the
$K=1$ situation when the carrying capacity is of ${\cal C}_1(x)$ type:
as illustrated in Fig.~\ref{fig:pddef1}, the new bottleneck
(sub-)phases are BP, MC-BP, LD-P and LD-BP-HD (see Fig.~\ref{fig:eg3}
(left) for an example of current and density profiles).  Again, we
notice that the core of the phase-diagram is entirely determined by
the defect (shadowed region in Fig.~\ref{fig:pddef1}), which is also
responsible for a pure BP for sufficiently high rates $\alpha$ and
$\beta$.

On the other hand, when $K\neq 1$ it follows from the analytical
solution of the above-mentioned MF equation that there is only one
screening length on the subsystem \r. In fact, it turns out that the
left branch of $j_d(x)$ can never reach the maximal current $j^*_K(x)$
on the subsystem \l (so, for $K\neq 1$, there is no finite screening
length on the subsystem \l): the screening length may only be shorter
than the length of the subsystem \r, with the point $x_{\d}+\xi<1$. It
follows that carrying capacities of types ${\cal C}_2(x)$ and ${\cal
  C}_4(x)$ are topologically prohibited (even in the asymmetric case
$x_{\d}\neq 1/2$) when $K\neq 1$.

In contrast to the case $K=1$, as in general the current $j_\beta(x)$
can not reach the value $j^*_K(x)$, the carrying capacity of type
${\cal C}_3(x)$ gives rise to four bottleneck phases (instead of six
as for the case $K=1$), namely the LD-BP-HD, LD-BP-MC, BP-HD and
BP-MC. An example of BP-MC current and density profiles is shown in
Fig.~\ref{fig:eg3} (right).

\section{Conclusion}
\label{sec:concl}
This work has been devoted to the study of the influence of a
bottleneck (point-wise disorder) on the stationary properties of a
biologically inspired stochastic transport model obtained by coupling
two paradigmatic equilibrium and non-equilibrium processes. Namely, we
have considered the competition between the \emph{totally asymmetric
  exclusion process} (TASEP) and Langmuir kinetics (LK) in the
presence of open boundaries and a bottleneck, which locally slows down
any incoming particles.  The current and density profiles in the
non-equilibrium steady state have been investigated analytically via
an effective mean-field theory built on splitting the lattice into two
subsystems.  Our analytical results were checked against numerical
(Monte-Carlo) simulations. 

As a consequence of the competition between the TASEP and LK dynamics,
the effects of a single bottleneck in the TASEP/LK model are much more
dramatic than in the simple TASEP \cite{kolomeisky:98}, or the
so-called $\ell-$TASEP (TASEP for extended objects)
\cite{shaw-kolomeisky-lee:03}, where a localized defect was shown to
merely shift some transitions line in the phase-diagram, but do not
affect its topology. Here, new and mixed phases induced by the
bottleneck have been obtained.

As a key concept of our analysis, we have introduced the {\it carrying
  capacity}, which is defined as the maximal current that can flow
through the bulk of the system. In contrast to the simple TASEP the
spatial dependence of the current, caused by the Langmuir kinetics,
makes the carrying capacity non-trivial: The defect depletes the
current profile within a distance that we called \emph{screening
  length}. This quantity increases with the strength of the defect and
decreases with the attachment-detachment rates. The competition
between the current imposed at the boundaries and the one limited by
the defect determines the density profiles and the ensuing phase
diagram.  When the boundary currents dominant, the phase behavior of
the defect-free system is recovered. Also, above some critical
entrance and exit rates, the system transports the maximal current,
independently of the boundaries.  Between these two extreme
situations, we have found several coexistence phases, where the
density profile exhibits stable shocks and kinks. Indeed, above some
specific parameter values the phase-diagram is characterized by
\emph{bottleneck phases}. Depending on the screening length imposed by
the defect, which can cover the entire system or part of it, different
phases-diagrams arise. The latter are characterized by four, six or
nine bottleneck phases, which have been quantitatively studied within
our mean-field theory.

The analysis carried out in this work can be straightforwardly
extended to many variants of the TASEP/LK model. As an example, let us
mention the case of a lattice gas where dimers (modelling the usual
two heads of molecular motors) would move as bound entities according
to the $\ell-$TASEP (with $\ell=2$) \cite{shaw-zia-lee:03} and could
experience an Langmuir-like on-off kinetics.  Recently, such a system
has been studied (without disorder) within an appropriate
mean-field-like scheme \cite{pierobon-franosch-frey:06}. The
point-wise version of this system could be investigated along the same
lines described in the present work.

In addition to direct extensions, we think that the method outlined in
this paper could pave the way to study the TASEP/LK models in more
`realistic' and biophysically relevant situations, as in the presence
of clusters of competing defects or quenched site-wise randomness.

\acknowledgments The authors are grateful to T. Reichenbach, A.
Parmeggiani and T. Franosch for useful discussions.  M.M. gratefully
acknowledges a fellowship of the German Alexander-von-Humboldt
Foundation (grant IV-SCZ/1119205 STP).


\begin{thebibliography}{41}
\bibitem{foot3}{For $q>1$ one expects only variations of the dynamics
    but not of the stationary properties; the phase diagram is the
    same as for the defect-free model.}
\end{thebibliography}
\end{document}